%% file: ctcomp.tex
\documentclass[11pt,osf]{econart}

\usepackage{enumerate}
\usepackage{algorithm}
\usepackage{algpseudocode}
\usepackage{booktabs}
\usepackage{subcaption}

\usepackage{tikz}
\usetikzlibrary{arrows.meta, positioning, automata, decorations.markings, calc, backgrounds, fit}

\def\Nplayers{N}
\def\Ndemand{D}
\def\Nchoices{J}
\def\choiceset{\mathcal{J}}
\def\vecstatespace{\mathcal{X}}
\def\intstatespace{\mathcal{K}}
\def\unifrate{\eta}

\def\accumv{w}
\DeclareMathOperator*{\argmax}{arg\,max}

\def\thetaD{\theta_{\text{D}}}
\def\thetaRN{\theta_{\text{RN}}}
\def\thetaEC{\theta_{\text{EC}}}
\def\nactivek{n_{k}}

\defcitealias{abbe-2016}{ABBE (2016)}
\shortcites{rupp-2024}

\def\allassumptions{Assumptions~\ref{assp:dx}--\ref{assp:pi}}

\renewcommand*{\backref}[1]{}

\begin{document}

\title{Leveraging Uniformization and Sparsity for Estimation and
Computation of Continuous-Time Dynamic Discrete Choice Games}

\author{\uppercase{Jason R. Blevins} \\ The Ohio State University}

\pdfauthor{Jason R. Blevins}


\date{November 7, 2025}


\abstract{%
Continuous-time empirical dynamic discrete choice games
offer notable computational advantages over discrete-time models.
This paper addresses remaining computational and econometric challenges
to further improve both model solution and estimation.
We establish convergence rates for value iteration and policy evaluation
with fixed beliefs, and develop Newton-Kantorovich
methods that exploit analytical Jacobians and sparse matrix structure.
We apply uniformization both to derive a new representation of the value function
that draws direct analogies to discrete-time models and to enable
stable computation of the matrix exponential and its parameter derivatives
for estimation with discrete-time snapshot data, a common but
challenging data scenario.
These methods provide a complete chain of analytical derivatives
from the value function for a given equilibrium through the log-likelihood function,
eliminating the need for numerical differentiation and improving
finite-sample estimation accuracy and computational efficiency.
Monte Carlo experiments demonstrate substantial gains in both statistical
performance and computational efficiency, enabling researchers to estimate
richer models of strategic interaction.
While we focus on games, our methods extend to single-agent dynamic
discrete choice and continuous-time Markov jump processes.}

\acknowledgments{%
  Python replication code is available at \url{https://github.com/jrblevin/ctcomp}.
  I am grateful to Victor Aguirregabiria, Adam Dearing,
  Terence Johnson, and Larry Samuelson as well as
  participants of the 2024 Empirical Methods in Game Theory Workshop
  at Stony Brook University, the 2025 IIOC Conference,
  and the 2025 Midwest Econometrics Group conference
  for helpful feedback and discussions.
}

\maketitle

\section{Introduction}
\label{sec:intro}

Understanding the dynamics of market structure and competition in
oligopolistic industries is a foundational goal of modern empirical
industrial organization.
Beginning with the seminal works of \cite{bresnahan91empirical}
and \cite{berry-1992}, the field has transitioned from static models of market
entry to dynamic models that capture the forward-looking behaviors of firms,
influenced by methodological advancements from \cite{aguirregabiria-mira-2007},
\cite{bajari-benkard-levin-2007}, \cite{pakes-ostrovsky-berry-2007},
and \cite{pesendorfer08asymptotic}.
However, this shift introduced significant computational challenges,
where solving discrete-time dynamic equilibrium models becomes exponentially
more complex as the number of players increases.
In discrete-time games with simultaneous moves, computing equilibria requires
forming expectations over all possible combinations of rivals' actions, and
the complexity grows exponentially with the number of players.
These computational barriers have limited empirical applications and
counterfactuals to relatively
simple models with few players and states, preventing researchers from
analyzing realistic market structures.
The prevalence of snapshot data---observations recorded at evenly spaced
intervals without exact transition times---has driven the focus on
discrete-time models with simultaneous moves in empirical work.
However, the underlying economic reality often involves sequential
decision-making---whether observed or not---with distinct strategic
implications.

The introduction of continuous-time dynamic discrete choice games
by \cite{doraszelski-judd-2012} and \citet*{abbe-2016} was aimed
at addressing the computational limitations of discrete-time models.
These models offer a more granular representation of decision-making
processes which unfold in a stochastic, sequential manner and align
with the nature of strategic interactions in many real-world settings.
\cite{doraszelski-judd-2012} showed that continuous-time modeling
simplifies the computation of players' expectations over future
states, thereby reducing the computational costs relative to
discrete-time models.
Because only one player moves at any instant in continuous time,
the order of operations required to compute equilibria scales linearly
with the number of players rather than exponentially.
\citet*{abbe-2016} developed an empirical framework and two-step
CCP estimator based on the seminal contributions of
\cite{hotz93conditional} and \cite{hotz94simulation}.
More recently, \cite{blevins-kim-2024} extended the nested pseudo
likelihood estimator of \cite{aguirregabiria-mira-2007} to the case
of continuous-time games and
\cite{ctgames} considered identification and estimation of
move arrival rates in the model, which were previously held fixed.

Despite these advancements, estimation of continuous-time models faces
computational challenges that have limited their adoption.
First, the convergence properties of value iteration had not been formally
established, leaving researchers without theoretical guarantees about
algorithm behavior.
Second, reliance on numerical derivatives for optimization introduces
approximation errors that propagate into parameter estimates, degrading
finite-sample performance.
Third, for researchers using snapshot data, computation of the matrix
exponential $\exp(\Delta Q)$ is necessary for likelihood evaluation
but is computationally intensive, particularly as model size grows.\footnote{
  In contrast, continuous-time data with observed
  transition times allow likelihood computation directly from the
  intensity matrix $Q$ without matrix exponentials
  \citep{billingsley61statistical}.}
The sparsity of $Q$ is a key feature our methods exploit for computational
efficiency in both equilibrium computation and matrix exponential calculation.
While the value function methods we develop apply regardless of data structure,
our matrix exponential algorithms specifically target the snapshot data case.

This paper provides theoretical foundations and
computational methods to enhance the solution and estimation of
continuous-time dynamic discrete choice models.
In Section~\ref{sec:model}, we review the model and assumptions.
In Section~\ref{sec:vf}, we consider computation of value functions and
equilibria.
We establish the contractivity and modulus of contraction of the
Bellman optimality operator when beliefs are held fixed, filling a gap
in the theoretical foundations for this class of models.
Building on this, we develop a polyalgorithm that combines
the global convergence properties of value iteration with the quadratic convergence
of Newton-Kantorovich methods.
This approach, inspired by the nested fixed point
(NFXP) algorithm of \cite{rust87optimal} but adapted for continuous-time settings,
switches algorithms based on convergence monitoring to achieve
both reliability and efficiency in solving the model.
\cite{iskhakov16comment} reinforce the importance of Newton-Kantorovich iterations
for the performance of NFXP in discrete-time models.
We then develop a new representation of the value function based on
the method of uniformization, introduced by \cite{jensen-1953}
for continuous-time Markov chain analysis,
which allows us to draw direct comparisons to the discrete-time dynamic
discrete choice literature and to establish rates of convergence for
policy evaluation.

In Section~\ref{sec:loglik}, we turn to computation of the log-likelihood
function with snapshot data.
We show how to apply uniformization to develop an
efficient method for simultaneous computation of the matrix exponential
and its derivatives in a way that is optimized
for the sparse structure of transition rate matrices in the model.
This is important for computationally efficient and accurate estimation
of the model's structural parameters.
Using analytical derivatives eliminates a source of approximation
error that affects not just computational efficiency but also the
finite sample properties of estimators in practice.

While we develop these methods in the context of continuous-time
dynamic games to address their specific computational challenges, the
techniques have broader applicability.
Our application of uniformization for computing matrix exponentials
and their derivatives extends to any finite-state continuous-time
Markov chain with sparse intensity matrices, making the approach
valuable for single-agent dynamic discrete choice models, duration
models with time-varying covariates, and other finite-state
continuous-time econometric models.

To demonstrate the practical utility of the methods developed, in
Section~\ref{sec:mc} we conduct a series of Monte Carlo experiments
using a dynamic entry/exit model with stochastic market demand.
These experiments focus on evaluating the computational performance
of our analytical gradient and sparse matrix exponential methods.
We find that analytical gradients reduce computation time by 46\% and function
evaluations by 84\% compared to numerical gradients, while sparse
matrix methods achieve speedups of up to 40$\times$ when computing only
the columns of the matrix exponential needed for likelihood evaluation,
rather than computing the full dense matrix.
By identifying these computational efficiencies, we facilitate the
estimation and simulation of larger, more complex dynamic discrete choice
models in continuous time, allowing for deeper insights into strategic
interactions and policy interventions in oligopolistic markets.

\section{Continuous-Time Dynamic Discrete Choice Games}
\label{sec:model}

\subsection{Structural Model}
\label{sec:model:structural}

We model strategic interactions in a continuous-time dynamic discrete
choice framework with $\Nplayers$ forward-looking agents, indexed by
$i = 1, \dots, \Nplayers$, who operate over an infinite time horizon
$t \in [0, \infty)$.  The model accommodates both the stochastic nature
of decision times and the dynamic considerations of agents making
decisions with persistent effects.

\paragraph{State Space}

At any instant $t$, the state of the system is represented by a state
vector $x \in \vecstatespace$ that includes all payoff-relevant
information that is common knowledge among the agents.
This vector typically includes both agent-specific attributes
(e.g., market entry status, product quality)
and exogenous market conditions (e.g., demand levels, available
technology) that affect decisions and payoffs.
The state space is finite, with $K = |\mathcal{X}|$
denoting the total number of states.
We will generally work with the integer state space
$\intstatespace = \{1, \dots, K\}$ and refer to states by their index
$k$ or as an indexed vector $x_k$, as this allows us to vectorize many
expressions.

\paragraph{Decisions and Endogenous State Changes}

Decision opportunities for agent $i$ in state $k$ arrive according to
independent Poisson processes with finite rate parameters $\lambda_{ik}$,
reflecting limited attention or other frictions on decision frequency.
At each decision time, the agent observes $x_k$ and chooses an
action $j$ from the choice set $\choiceset_{ik}$.
For expositional clarity, we focus on a common choice set:
$\choiceset_{ik} = \choiceset = \{0, 1, \ldots, \Nchoices-1\}$.
The outcome of agent $i$'s decision $j$ in state $k$ is a
deterministic state transition to $k' = l(i,j,k)$.

\paragraph{Exogenous State Changes}

An artificial player called ``nature'' ($i = 0$) governs exogenous
state transitions, which follow a Markov jump process characterized
by an intensity matrix $Q_0$.
The $(k,k')$ element $q_{0kk'}$ represents the rate at which the
system transitions from state $k$ to state $k'$ due to exogenous
factors, with the aggregate exit rate from state $k$ given by
$\sum_{k' \neq k} q_{0kk'}$.

\paragraph{Payoffs}

Agent $i$ receives flow payoffs $u_{ik}$ while in state $k$, discounted
at rate $\rho_i$, with present value
$\int_0^\tau e^{-\rho_i t} u_{ik}\,dt = u_{ik} \left(1 - e^{-\rho_i \tau} \right)/\rho_i$
over interval $[0,\tau)$.
In contrast, instantaneous choice-specific payoffs $c_{ijk}$ are incurred
when agent $i$ chooses action $j$ in state $k$.
These payoffs are additively separable:
\begin{equation*}
c_{ijk} = \psi_{ijk} + \varepsilon_{ijk},
\end{equation*}
where $\psi_{ijk}$ is the deterministic mean payoff and
$\varepsilon_{ijk}$ is an unobserved choice-specific shock.
We normalize $\psi_{i0k} = 0$ for the continuation action $j=0$.

\begin{example}[Two-Player Entry/Exit Model]
\label{ex:entry}
In a market entry and exit game with $\Nplayers = 2$ firms and
$\Ndemand = 2$ demand states $\text{L}$ and $\text{H}$, each state
is described by $x_k = (x_{k0}, x_{k1}, x_{k2})$ where
$x_{k0} \in \{\text{L}, \text{H}\}$ is the demand level
(controlled by nature, $i=0$)
and $x_{ki} \in \{0,1\}$ is the activity indicator for firm $i \in \{1,2\}$.
The state space is
\begin{equation*}
  \begin{array}{rlllll}
    \vecstatespace = \{ & (\text{L}, 0, 0), & (\text{L}, 1, 0), & (\text{L}, 0, 1), & (\text{L}, 1, 1), & \\
                        & (\text{H}, 0, 0), & (\text{H}, 1, 0), & (\text{H}, 0, 1), & (\text{H}, 1, 1) & \}, \\
  \end{array}
\end{equation*}
giving us $K = 8$ total states that can be equivalently represented as
$\intstatespace = \{ 1, \dots, 8 \}$.

Each firm has action space $\choiceset = \{0, 1\}$ where
$j = 0$ represents \emph{maintaining current status} and
$j = 1$ represents \emph{switching market status}.
Action $j=0$ maintains the current state, $l(i, 0, k) = k$, while
action $j=1$ switches firm $i$'s status, $l(i, 1, k) = k'$, where
$x_{k'}$ differs from $x_k$ only in the activity indicator for firm $i$.
Note that firms cannot directly change the demand state, which is
controlled by nature.
Suppose demand transitions between states $\text{H}$ and $\text{L}$
at rate $\gamma$ in both directions.
Then,
\begin{equation*}
  q_{0kk'} = \begin{cases}
    \gamma & \text{for } (k,k') \in \{ (1,5), (2,6), (3,7), (4,8) \}, \\
    \gamma & \text{for } (k,k') \in \{ (5,1), (6,2), (7,3), (8,4) \}, \\
    0 & \text{otherwise}.
  \end{cases}
\end{equation*}

The flow payoffs in each state depend on
both competition (choices of rival firms) and demand.
Here $x_{k0}$ is the demand state and
$x_{ki}$ is the activity indicator for firm $i$ in state $k$.
Let $\nactivek = \sum_{i = 1}^\Nplayers x_{ki}$ denote the number of active
firms in state $k$.
Then the payoff for firm $i$ in state $k$ is given by
\begin{equation*}
u_{ik} = x_{ki} \left(\thetaRN \nactivek + \thetaD \1\{ x_{k0} = \text{H}\} \right),
\end{equation*}
where $\thetaRN$ is the competitive effect and $\thetaD$ is the demand effect.
Entry incurs a cost $\thetaEC < 0$, while exit is costless, so the instantaneous
payoffs are
\begin{equation*}
\psi_{ijk} = \begin{cases}
  \thetaEC & \text{if } j = 1 \text{ and } x_{ki} = 0, \\
  0 & \text{otherwise}.
\end{cases}
\end{equation*}
\end{example}

\paragraph{Assumptions}

Before describing how agents determine their actions in equilibrium,
we formalize the main assumptions of the model, following
\citet*{abbe-2016} and \cite{ctgames}:

\begin{assumption}[Discrete States]
  \label{assp:dx}
  The state space is finite:
  $K \equiv \abs{\mathcal{X}} < \infty$.
\end{assumption}

\begin{assumption}[Discount Rates]
  \label{assp:rho}
  Discount rates $\rho_i \in (0,\infty), i = 1, \dots, \Nplayers$ are known.
\end{assumption}

\begin{assumption}[Bounded Rates]
  \label{assp:rates}
  Rates for decision times and exogenous state changes satisfy
  $0 \leq \lambda_{ik} < \infty$, $0 \leq q_{0kk'} < \infty$,
  and $\sum_{k' \neq k} q_{0kk'} + \sum_m \lambda_{mk} > 0$
  for all $i = 1, \dots, \Nplayers$, and $k, k' \in \intstatespace$.
\end{assumption}

\begin{assumption}[Bounded Payoffs]
  \label{assp:u}
  Flow payoffs and choice-specific payoffs satisfy
  $\abs{u_{ik}} < \infty$ and $\abs{\psi_{ijk}} < \infty$ for all
  $i = 1, \dots, \Nplayers$, $j \in \choiceset$, and $k \in \intstatespace$.
\end{assumption}

\begin{assumption}[Additive Separability]
  \label{assp:as}
  Instantaneous payoffs are additively separable as
  $c_{ijk} = \psi_{ijk} + \varepsilon_{ijk}$.
\end{assumption}

\begin{assumption}[Costless Continuation \& Distinct Actions]
  \label{assp:dn} 
  For all $i = 1, \dots, \Nplayers$ and $k \in \intstatespace$:
  (a) $l(i,0,k) = k$ and $\psi_{i0k} = 0$, and
  (b) $l(i,j,k) \neq l(i,j',k)$ for all $j, j' \in \choiceset$ with $j' \neq j$.
\end{assumption}

\begin{assumption}[Private Information]
  \label{assp:pi}
  Choice-specific errors
  $\varepsilon_{ik} = (\varepsilon_{i0k}, \ldots, \varepsilon_{i,\Nchoices-1,k})^\top$
  are iid across players, states, and decision times with common
  distribution $F$.
  The distribution $F$ has a continuous density with respect to
  Lebesgue measure, has finite first moment, and has support $\R^{\Nchoices}$.
\end{assumption}

The previously mentioned Assumptions~\ref{assp:dx}--\ref{assp:as} lay
the groundwork for the model's structure and dynamics.
Assumption~\ref{assp:dn} clarifies the role of the action $j = 0$,
designating it as a continuation action that does not alter the current
state.
It also stipulates actions $j > 0$ are meaningfully different from one
another, for identification purposes.
Assumption~\ref{assp:pi} describes our assumptions on the
idiosyncratic error terms in the model and mirrors similar assumptions
used in discrete time models \citep{aguirregabiria-mira-2010}.

\paragraph{Strategies, Beliefs, and Value Functions}

A stationary Markov strategy for agent $i$, denoted $\delta_i$, maps
each state $k \in \intstatespace$ and vector of choice-specific shocks
$\varepsilon_{ik} \in \R^\Nchoices$ to an action in $\choiceset$.
Any strategy $\delta_i$ determines choice probabilities
\begin{equation}
  \label{ccps}
  \sigma_{ijk} = \Pr\left[ \delta_i(k, \varepsilon_{ik}) = j \mid k \right]
\end{equation}
for all choices $j$ and states $k$.

To determine an optimal strategy, player $i$ must form beliefs about the
choice probabilities of rivals.
Let $\varsigma_{imjk}$ denote player $i$'s belief that rival $m$ chooses
action $j$ in state $k$, with
$\varsigma_{im} = \left(\varsigma_{imjk}\right)_{j\in\choiceset,k\in\intstatespace}$
denoting the beliefs about rival $m$, and
\begin{equation}
  \label{eq:beliefs}
  \varsigma_i = (\varsigma_{i1}, \dots, \varsigma_{i,i-1}, \varsigma_{i,i+1}, \dots, \varsigma_{i\Nplayers})
\end{equation}
denoting player $i$'s complete beliefs about all rivals.

Given beliefs $\varsigma_i$, the value function for player $i$ is
$V_i^{\varsigma_i} = (V_{i1}^{\varsigma_i}, \ldots, V_{iK}^{\varsigma_i})^\top$
where each element represents the expected present value of future
rewards beginning in state $k$ and making optimal decisions
thereafter.
Following \citet[][Section 2.7]{ctgames},
the Bellman equation for $V_{ik}^{\varsigma_i}$ is:
\begin{small}
\begin{equation}
  \label{eq:bellman}
  V_{ik}^{\varsigma_i} = \frac{u_{ik} +
    \sum_{k' \neq k} q_{0kk'} V_{ik'}^{\varsigma_i} +
    \sum_{m \neq i} \lambda_{mk} \sum_j \varsigma_{imjk} V_{i,l(m,j,k)}^{\varsigma_i} +
    \lambda_{ik} \E \max_j \{ \psi_{ijk} + \varepsilon_{ijk} + V_{i,l(i,j,k)}^{\varsigma_i} \}}
  {\rho_i + \sum_{k' \neq k} q_{0kk'} + \sum_{m=1}^{\Nplayers} \lambda_{mk}}.
\end{equation}
\end{small}

\noindent
This representation shows the balance between the immediate flow
payoff in state $k$, the expected values following exogenous and
endogenous state transitions, and the anticipated outcomes of player
$i$'s decisions, all adjusted for the probability of occurrence
and discounted appropriately.
The expectation is over the joint distribution of the vector of shocks
$\varepsilon_{ik}$.
The denominator includes the total rate of events that can occur in state $k$,
which we refer to as the \emph{maximum exit rate} from state $k$, denoted as
\begin{equation}
  \label{eq:unifrate_k}
  \unifrate_k \equiv \sum_{k' \neq k} q_{0kk'} + \sum_{m=1}^{\Nplayers} \lambda_{mk}.
\end{equation}

\begin{definition}
The \emph{Bellman optimality operator} $T_{i}^{\varsigma_i}$ for player $i$
with beliefs $\varsigma_i$ is defined by stacking \eqref{eq:bellman} across
states $k = 1, \dots, K$.
The value function $V_i^{\varsigma_i}$ is a fixed point satisfying
$V_i^{\varsigma_i} = T_{i}^{\varsigma_i} V_i^{\varsigma_i}$.
\end{definition}

\paragraph{Markov Perfect Equilibrium}

We now define the equilibrium concept used in this paper.  A profile
of stationary Markov strategies
$\delta = (\delta_1, \ldots, \delta_\Nplayers)$ determines a choice
probability profile $\sigma = (\sigma_1, \ldots, \sigma_\Nplayers)$,
where $\sigma_i = \{\sigma_{ijk}\}_{j,k}$.
A strategy $\delta_i$ is a best response to beliefs $\varsigma_i$
if it assigns action $j$ that maximizes expected discounted utility
in state $k$:
\begin{equation}
  \label{prule}
  \delta_i(k,\varepsilon_{ik}) = j
  \iff
  \psi_{ijk} + \varepsilon_{ijk} + V_{i,l(i,j,k)}^{\varsigma_i}
  \geq
  \psi_{ij'k} + \varepsilon_{ij'k} + V_{i,l(i,j',k)}^{\varsigma_i}
  \quad \forall j' \in \choiceset.
\end{equation}

\begin{definition}[Markov Perfect Equilibrium]
A Markov perfect equilibrium (MPE) is a profile of choice probabilities
$\sigma^* = (\sigma_1^*, \ldots, \sigma_\Nplayers^*)$ such that for each player $i$:
\begin{enumerate}
\item Player $i$'s choices are optimal given beliefs $\varsigma_i^*$:
  \begin{equation*}
    \sigma_{ijk}^* = \Pr\left[ j \in \argmax_{j' \in \choiceset} \left\{ \psi_{ij'k} + \varepsilon_{ij'k} + V_{i,l(i,j',k)}^{\varsigma_i^*} \right\} \right] \quad\text{for all } j, k,
  \end{equation*}
where $V_i^{\varsigma_i^*}$ is a fixed point of the Bellman operator $T_i^{\varsigma_i^*}$.
\item Player $i$'s beliefs $\varsigma_i^*$ are consistent:
  $\varsigma_{imjk}^* = \sigma_{mjk}^*$ for all $m \neq i$, $j$, and $k$,
\end{enumerate}
\end{definition}

We can characterize equilibria of the model as fixed points of a
nonlinear, simultaneous system of equations for the value functions $T(V) = V$,
where the system operator $T: \R^{\Nplayers \times K} \rightarrow \R^{\Nplayers \times K}$
applies each player's Bellman optimality operator given the beliefs
$\varsigma_i(V_{-i})$ implied by the value functions of other players $V_{-i}$:
\begin{equation}
  \label{eq:equilibrium_system}
  T(V) = \begin{bmatrix}
    T_1^{\varsigma_1(V_{-1})}(V_1) \\
    T_2^{\varsigma_2(V_{-2})}(V_2) \\
    \vdots \\
    T_\Nplayers^{\varsigma_\Nplayers(V_{-\Nplayers})}(V_\Nplayers) \\
  \end{bmatrix}.
\end{equation}

Existence of MPE was established by
\citet*{abbe-2016} in models with homogeneous rates and by
\cite{ctgames} in the present model with heterogeneous rates.

\subsection{Continuous-Time Markov Chains and Uniformization}
\label{sec:model:ctmc}

The model's dynamics follow a finite-state continuous-time Markov chain
(CTMC), or Markov jump process, denoted $X(t)$ for
$t \in [0, \infty)$ with values in the state space $\mathcal{X}$.
\footnote{We refer the reader to
\citet[Section 4.8]{karlin75first},
\citet[Chapter 4]{tijms-2003},
or \citet[part II]{chung-1967} for details.}
We characterize such a process using its $K \times K$ \emph{intensity matrix}
$Q = (q_{kk'})$ where, for $k' \neq k$,
\begin{equation*}
  q_{kk'} = \lim_{h \to 0} \frac{\Pr\left[X(t+h) = k' \mid X(t) = k \right]}{h}
\end{equation*}
represents the instantaneous rate at which transitions occur from
state $k$ to state $k'$.  The diagonal elements, $q_{kk}$, are set to
$-\sum_{k' \neq k} q_{kk'}$, ensuring row sums are zero.
The rate parameter for the exponential distribution of holding times
before exiting state $k$ is $-q_{kk}$, equal to the aggregate
off-diagonal transition rates out of state $k$.
When the process exits state $k$, it transitions to state
$k' \neq k$ with probability $q_{kk'} / (-q_{kk})$.

For estimation with snapshot data sampled at intervals of length
$\Delta$, we require the \emph{transition probability matrix}
$P(\Delta) = \exp(\Delta Q)$, whose $(k,k')$ element gives the
probability of transitioning from state $k$ to state $k'$ over
time $\Delta$.
The matrix exponential is defined by its power
series expansion:
\begin{equation}
  \label{eq:expm}
  P(\Delta)
  = \exp(\Delta Q)
  = \sum_{j=0}^\infty \frac{(\Delta Q)^j}{j!}
  = I + \Delta Q + \frac{(\Delta Q)^2}{2!} + \frac{(\Delta Q)^3}{3!} + \cdots.
\end{equation}

Despite its theoretical appeal, directly calculating the matrix
exponential via \eqref{eq:expm} is problematic.
Matrix powers can be numerically unstable, and catastrophic
cancellation may occur from alternating signs since $Q$ contains both
positive and negative elements.
Furthermore, the direct approach fails to exploit sparsity,
as matrix powers of sparse $Q$ become dense as $j$ increases.
While various numerical methods have been developed to address these issues,
including Pad\'{e} approximations and scaling-and-squaring methods
\citep{moler78nineteen,moler-van-loan-2003}, we employ uniformization
\citep{jensen-1953}, a technique tailored specifically
for generator matrices of continuous-time Markov chains.
This approach naturally exploits
sparsity when computing $\exp(\Delta Q) v$ for vectors $v$---a
central feature of the models we consider.\footnote{Other contributions to
uniformization include \cite{grassmann-1977-ejor, grassmann-1977-compor},
\cite{reibman-trivedi-1988}, and \cite{sherlock-2022}.
See \cite{vandijk-2018-uniformization} for a recent overview.}

Uniformization converts a continuous-time Markov chain with varying
exit rates into a discrete-time Markov chain subordinate to a Poisson
process with rate parameter $\unifrate$ satisfying
\begin{equation}
  \label{eq:unifrate}
  \unifrate \geq \max_{k\in\intstatespace} \abs{q_{kk}}.
\end{equation}
This is achieved by introducing self-transitions that allow the process
to remain in its current state.
The discrete-time transition probability matrix is constructed as:
\begin{equation*}
  \Sigma = I + \frac{Q}{\unifrate}.
\end{equation*}
The probability of transitioning from state $k$ to state $k'$ in time
$\Delta$ is then given by
\begin{equation}
  \label{eq:expm:unif}
  P(\Delta)_{kk'} = e^{-\unifrate \Delta} \sum_{j=0}^{\infty} \frac{(\unifrate \Delta)^j}{j!} [\Sigma^j]_{kk'},
\end{equation}
where $[\Sigma^j]_{kk'}$ is the probability of transitioning from $k$ to $k'$
in exactly $j$ jumps, which is weighted by the Poisson probability of $j$
arrivals within interval $\Delta$.
While any $\unifrate$ satisfying \eqref{eq:unifrate} is valid, in practice
using the smallest valid rate $\unifrate = \max_{k} \abs{q_{kk}}$ minimizes
the number of terms required in \eqref{eq:expm:unif} to achieve a given
accuracy (determined by the Poisson parameter $\unifrate \Delta$).

\subsection{Uniformization of the Two-Firm Entry/Exit Model}
\label{sec:model:example:entry:unif}

We now illustrate the construction and uniformization of equilibrium
intensity matrices using the market entry and exit model introduced
earlier.
Recall that with $\Nplayers = 2$ firms and 2 demand states, we have $K = 8$
total model states.
The aggregate intensity matrix $Q$ decomposes as $Q = Q_0 + Q_1 + Q_2$, where
$Q_0$ captures exogenous demand transitions and $Q_i$ captures firm $i$'s
endogenous decisions.

Demand switches between H and L at rate $\gamma$, so $Q_0$ has the block form:
\begin{equation*}
Q_0 = \begin{bmatrix}
-\gamma I_4 & \gamma I_4 \\
\gamma I_4 & -\gamma I_4
\end{bmatrix},
\end{equation*}
where $I_4$ is the $4 \times 4$ identity matrix.
This structure reflects the fact that exogenous demand changes
do not change firm activity configurations.

For simplicity, assume that firms make decisions at the same rate across
states so that $\lambda_{ik} = \lambda$ for all $i$ and $k$.
Then firm 1's entry and exit decisions create transitions at rate
$\lambda \sigma_{11k}$:
\begin{equation*}
Q_1 = \begin{bmatrix}
-\lambda \sigma_{111} & \lambda \sigma_{111} & \cdots & 0 & 0 \\
\lambda \sigma_{112} & -\lambda \sigma_{112} & \cdots & 0 & 0 \\
\vdots & \vdots & \ddots & \vdots & \vdots \\
0 & 0 & \cdots & -\lambda \sigma_{117} & \lambda \sigma_{117} \\
0 & 0 & \cdots & \lambda \sigma_{118} & -\lambda \sigma_{118}
\end{bmatrix}.
\end{equation*}
Similarly, $Q_2$ captures firm 2's decisions with a different sparsity pattern.

The $8 \times 8$ aggregate intensity matrix (with diagonal elements omitted for brevity) is
\begin{equation}
\label{eq:entry:Q}
Q = \begin{bmatrix}
\cdot & \lambda\sigma_{111} & \lambda\sigma_{211} & 0 & \gamma & 0 & 0 & 0 \\
\lambda\sigma_{112} & \cdot & 0 & \lambda\sigma_{212} & 0 & \gamma & 0 & 0 \\
\lambda\sigma_{213} & 0 & \cdot & \lambda\sigma_{113} & 0 & 0 & \gamma & 0 \\
0 & \lambda\sigma_{214} & \lambda\sigma_{114} & \cdot & 0 & 0 & 0 & \gamma \\
\gamma & 0 & 0 & 0 & \cdot & \lambda\sigma_{115} & \lambda\sigma_{215} & 0 \\
0 & \gamma & 0 & 0 & \lambda\sigma_{116} & \cdot & 0 & \lambda\sigma_{216} \\
0 & 0 & \gamma & 0 & \lambda\sigma_{217} & 0 & \cdot & \lambda\sigma_{117} \\
0 & 0 & 0 & \gamma & 0 & \lambda\sigma_{218} & \lambda\sigma_{118} & \cdot
\end{bmatrix}.
\end{equation}
The aggregate $Q$ matrix exhibits significant sparsity because direct
transitions are only possible between states that differ in exactly
one component (either demand or one firm's entry status).
In this example, each row has at most 4 non-zero elements, yielding
50\% sparsity.
More generally, for $\Nplayers$ firms and $\Ndemand$ demand states,
each state connects to at most $\Nplayers + \Ndemand$ other states,
while the total state space has $2^{\Nplayers} \times \Ndemand$ elements.
As we show later in Table~\ref{tab:sparsity},
with $5$ players and $3$ demand states ($K = 96$ total states),
the $Q$ matrix is over 90\% sparse, and the sparsity ratio approaches 1
as $\Nplayers$ grows, making sparse matrix algorithms essential for
computational tractability in realistic applications.

\definecolor{firmOneColor}{RGB}{0, 0, 255}         
\definecolor{firmTwoColor}{RGB}{34, 139, 34}       
\definecolor{demandColor}{RGB}{180, 20, 60}        
\definecolor{stateColor}{RGB}{255, 255, 255}       
\definecolor{stateBorderColor}{RGB}{0, 0, 0}       
\definecolor{groupColor}{RGB}{248, 248, 248}       

\newcommand{\firmOneText}[1]{\textcolor{firmOneColor}{#1}}
\newcommand{\firmTwoText}[1]{\textcolor{firmTwoColor}{#1}}
\newcommand{\demandText}[1]{\textcolor{demandColor}{#1}}

\begin{figure}[tbhp]
\begin{subfigure}{\textwidth}
\centering
\include{2x2x2-entry-standard.tex}
\caption{Continuous-time Markov chain $Q$}
\label{fig:entry:unif:standard}
\end{subfigure}
\\[3ex]
\begin{subfigure}{\textwidth}
\centering
\include{2x2x2-entry-uniform.tex}
\caption{Discrete-time Markov chain $\Sigma$ with self transition probabilities
$\Sigma_{kk} = 1 - \frac{\gamma + \lambda \sigma_{11k} + \lambda \sigma_{21k}}{\unifrate}$,
subordinate to a Poisson process with rate $\unifrate$}
\label{fig:entry:unif:uniform}
\end{subfigure}
\caption{Uniformization of the Two-Firm Entry/Exit Model}
\label{fig:entry:unif}
\end{figure}

To apply the uniformization technique, we select a uniform rate
$\unifrate$ satisfying \eqref{eq:unifrate} for any equilibrium $\sigma$:
$\unifrate = 2\lambda + \gamma$.
Given the $Q$ matrix in \eqref{eq:entry:Q},
the uniform transition probability matrix
$\Sigma = I + Q/\unifrate$ inherits the same sparsity pattern
and has the following structure.
For firm $i$ switching from state $k$ to $k'$,
where states $k$ and $k'$ differ only in firm $i$'s activity status,
the transition probabilities are
$\lambda \sigma_{ijk}/\unifrate$.
For demand transitions, where states $k$ and $k'$ have
the same firm configuration but demand levels that differ
by one unit, the transition probabilities are
$\gamma / \unifrate$.
Finally, the self-transition probabilities ensure that each row of
$\Sigma$ sums to one are
$\Sigma_{kk} = 1 - (\gamma + \lambda\sigma_{11k} + \lambda\sigma_{21k})/\unifrate$.

Figure~\ref{fig:entry:unif} illustrates this transformation, with panel
\subref{fig:entry:unif:standard} showing the original continuous-time chain and
panel \subref{fig:entry:unif:uniform} showing the uniformized discrete-time
chain. States are grouped by demand level, with horizontal transitions
representing firm 1's decisions and vertical transitions representing firm 2's
decisions.  Transitions between demand states for the same firm configuration
each occur with the same rate and are represented by single arrows.
The uniformization preserves the original dynamics while enabling
more efficient computation.

\section{Computing the Value Functions}
\label{sec:vf}

This section develops computational methods for solving continuous-time dynamic discrete choice games.
We first establish contraction properties of the Bellman operator and introduce a polyalgorithm
that combines value iteration with Newton-Kantorovich iterations for efficient equilibrium computation.
We then present a novel uniform policy-evaluation representation of the value function that connects continuous-time
models to discrete-time frameworks.

\subsection{Value Iteration}
\label{sec:vf:non-uniform}

We establish that the Bellman optimality operator $T_{i}^{\varsigma_i}$
defined in \eqref{eq:bellman}, holding beliefs $\varsigma_i$ fixed,
is a contraction with respect to the supremum norm.
While analogous results are well-known for discrete-time dynamic
programming, formal convergence guarantees have not been established
for the continuous-time models considered here.

\begin{theorem}[Value Iteration]
  \label{thm:vi:non-uniform}
  Suppose \allassumptions{} hold.
  For any player $i$ with fixed beliefs $\varsigma_i$,
  $T_{i}^{\varsigma_i}$ is a contraction with respect to the sup norm
  with modulus
  \begin{equation}
    \label{eq:beta_i}
    \beta_i \equiv \max_k \frac{\unifrate_k}{\rho_i + \unifrate_k}.
  \end{equation}
  Let $V_i^{(n)} \equiv \left(T_{i}^{\varsigma_i}\right)^{n} V_i^{(0)}$
  denote the $n$-th iterate of value iteration starting from any
  $V_i^{(0)} \in \R^K$.
  Then $\lim_{n \to \infty} V_i^{(n)} = V_{i}^{\varsigma_i}$,
  where $V_{i}^{\varsigma_i}$ is the unique fixed point of
  $T_{i}^{\varsigma_i}$, with
  \begin{equation*}
    \norm{V_i^{(n+1)} - V_{i}^{\varsigma_i}}_\infty \leq \beta_i \norm{V_i^{(n)} - V_{i}^{\varsigma_i}}_\infty
  \end{equation*}
  and convergence rate $O(\beta_i^n)$.
\end{theorem}

The proof of this theorem, and all remaining results,
is given in Appendix~\ref{sec:proofs}.

\begin{remark}
Theorem~\ref{thm:vi:non-uniform} and other results hold for single agent models
with $\Nplayers=1$ as a special case, where the beliefs
$\varsigma_i$ are not necessary as there are no rival players to consider.
\end{remark}

\subsection{Solving the Equilibrium System}
\label{sec:vf:equilibrium}

While Theorem~\ref{thm:vi:non-uniform} established contractivity of each
player's Bellman operator $T_{i}^{\varsigma_i}$ for fixed beliefs,
the equilibrium system operator $T(V)$ may not be globally contractive
because beliefs $\varsigma_i(V_{-i})$ are updated simultaneously with
value functions.
This makes equilibrium computation challenging, since multiple equilibria
may exist.\footnote{%
  Equilibrium multiplicity is well-recognized in static and dynamic games.
  See \cite{depaula-2013}, \cite{borkovsky-2015},
  \cite{aguirregabiria-collardwexler-ryan-2021}, and
  \cite{otsu-pesendorfer-2023}.}
Therefore, we focus on methods for efficiently computing \emph{an} equilibrium,
and we consider three approaches:
value iteration for finding equilibria that are stable under best-response dynamics,
Newton-Kantorovich iterations for finding regular equilibria with good initialization,
and a polyalgorithm that combines both methods.
The polyalgorithm, inspired by \citet{rust87optimal}'s approach for single-agent problems,
leverages the global properties of value iteration with the rapid local convergence
of Newton-Kantorovich iterations.
We discuss each method in turn.

\subsubsection{Value Iteration for the Equilibrium System}

Starting from any initial value function $V^{(0)}$,
value iteration updates all players' value functions simultaneously:
$V^{(n+1)} = T(V^{(n)})$.
Without global contractivity, multiple equilibria may arise and value
iteration alone can be slow to converge, particularly when the
contraction modulus is close to one.
Nevertheless, value iteration remains useful for finding equilibria that
are stable under the best-response dynamics implicit in $T(V)$.
An equilibrium $V^*$ is stable in this sense if the spectral radius of
$\frac{\partial T}{\partial V}(V^*)$ is less than unity, ensuring
that value iteration starting near $V^*$ will converge to $V^*$.
Unstable equilibria---where the spectral radius exceeds unity---will
have small or empty basins of attraction and will not be found unless
initialization is extremely close to them.

Following \cite[][p.~28]{nfxp}, we monitor the convergence rate
\begin{equation*}
r^{(n)} = \norm{V^{(n)} - V^{(n-1)}}_\infty / \norm{V^{(n-1)} - V^{(n-2)}}_\infty
\end{equation*}
to assess progress.
In practice, we take the target modulus to be $\beta = \max_i \beta_i$
where $\beta_i$ is defined in \eqref{eq:beta_i}.
When $r^{(n)}$ approaches $\beta$, the iteration is converging linearly at
the expected rate, suggesting that switching to Newton-Kantorovich iterations,
discussed in the next section, may be beneficial.

\subsubsection{Newton-Kantorovich Iterations}

While value iteration provides convergence to stable equilibria, it can be
slow when the contraction modulus is close to one.
On the other hand, Newton-Kantorovich iterations exhibit quadratic local convergence,
but rely on good initialization.
The method solves the fixed-point equation $V = T(V)$ by reformulating the problem
as finding a zero of $F(V) \equiv V - T(V)$ and applying Newton's method:
\begin{equation*}
  V^{(n+1)} = V^{(n)} - \left[I - \frac{\partial T}{\partial V}(V^{(n)})\right]^{-1} [V^{(n)} - T(V^{(n)})]
\end{equation*}

\citet{doraszelski-escobar-2010} show that regular Markov perfect
equilibria---where the Jacobian $I - \frac{\partial T}{\partial V}(V^*)$
is nonsingular---are generic in discrete-time dynamic stochastic games
with finite state and action spaces.
Nonsingularity ensures local uniqueness via the Inverse Function Theorem
and guarantees well-defined basins of attraction for Newton-Kantorovich
iterations \citep{ortega-rheinboldt-1970}.

To systematically search for unstable equilibria when needed, one can use
Newton-Kantorovich iterations directly with varied initial guesses.
The analytical Jacobian we derive can also serve to efficiently implement
alternative equilibrium-finding approaches, such as homotopy methods
\citep{besanko10learning,borkovsky10homotopy}, which rely on repeated
Jacobian evaluations along a continuation path.\footnote{%
  Other approaches for finding multiple equilibria include
  recursive lexicographical search \citep{iskhakov16recursive}, which requires
  additional structure on the state space and action sets.
  Two-step estimation methods
  can mitigate issues of multiple equilibria,
  bypassing explicit equilibrium computation \citep{blevins-kim-2024}.
}

\begin{table}[tb]
\centering
\caption{Structural Matrix Sparsity}
\label{tab:sparsity}
\begin{tabular}{rrrrrr}
\toprule
 &  & \multicolumn{2}{c}{Intensity Matrix $Q$} & \multicolumn{2}{c}{Jacobian $\partial T/\partial V$} \\
\cmidrule(lr){3-4} \cmidrule(lr){5-6}
Model & States ($K$) & Size ($K^2$) & Sparsity & Size ($K^2 \Nplayers^2$) & Sparsity \\
\midrule
2 $\times$ 2 & 8 & 64 & 50.00\% & 256 & 62.50\% \\
3 $\times$ 2 & 16 & 256 & 68.75\% & 2,304 & 81.25\% \\
4 $\times$ 2 & 32 & 1,024 & 81.25\% & 16,384 & 90.62\% \\
4 $\times$ 3 & 48 & 2,304 & 86.81\% & 36,864 & 93.58\% \\
5 $\times$ 3 & 96 & 9,216 & 92.36\% & 230,400 & 96.81\% \\
6 $\times$ 3 & 192 & 36,864 & 95.66\% & 1,327,104 & 98.41\% \\
6 $\times$ 4 & 256 & 65,536 & 96.68\% & 2,359,296 & 98.80\% \\
7 $\times$ 4 & 512 & 262,144 & 98.14\% & 12,845,056 & 99.40\% \\
7 $\times$ 5 & 640 & 409,600 & 98.50\% & 20,070,400 & 99.52\% \\
8 $\times$ 4 & 1,024 & 1,048,576 & 98.97\% & 67,108,864 & 99.70\% \\
8 $\times$ 5 & 1,280 & 1,638,400 & 99.17\% & 104,857,600 & 99.76\% \\
8 $\times$ 6 & 1,536 & 2,359,296 & 99.31\% & 150,994,944 & 99.80\% \\
9 $\times$ 5 & 2,560 & 6,553,600 & 99.55\% & 530,841,600 & 99.88\% \\
9 $\times$ 6 & 3,072 & 9,437,184 & 99.62\% & 764,411,904 & 99.90\% \\
10 $\times$ 6 & 6,144 & 37,748,736 & 99.79\% & 3,774,873,600 & 99.95\% \\
\bottomrule
\end{tabular}
\vspace{1ex}\newline
\footnotesize
Models are specified as $\Nplayers \times \Ndemand$, where $\Nplayers$ is the
number of players and $\Ndemand$ is the number of demand states.
Matrix size denotes the total number of elements.
Sparsity denotes the percentage of zero elements.
\end{table}

The key features that make Newton-Kantorovich tractable here
are the reduced computational burden of applying the Bellman operator
(linear complexity in $N$ rather than exponential)
and the sparse structure of continuous-time models.
Adapting this approach to multi-agent settings requires solving a
potentially large linear system of equations and computing the analytical
Jacobian of the equilibrium system---operations that would be computationally
prohibitive for discrete-time games of comparable size but remain tractable in
continuous time due to both the efficiency of computing expectations over
future states and the sparsity of the Jacobian.

Implementing the Jacobian $\frac{\partial T}{\partial V}$ efficiently
is important.
We compute the Jacobian analytically by
differentiating the equilibrium operator, with explicit formulas
derived in Appendix~\ref{sec:jacobian_structure}.
Importantly, both the Jacobian and the system matrix
$I - \frac{\partial T}{\partial V}$
inherit the sparsity of $Q$,
enabling the use of efficient sparse linear solvers such as GMRES
\citep{saad-schultz-1986}.
Table~\ref{tab:sparsity} illustrates this sparsity for the entry/exit model
from Section~\ref{sec:model:example:entry:unif}, reporting state space size
and sparsity of both $Q$ and $\frac{\partial T}{\partial V}$ across
several model sizes ($\Nplayers$ players and $\Ndemand$ demand states).
In models with 96 or more states, over 90\% of the elements
of both matrices are zeros.

The memory savings from sparse storage are substantial.
For dense storage, the Jacobian requires $O(K^2 \Nplayers^2)$ memory,
which becomes prohibitive for larger models.  For example, in the
$10 \times 6$ model ($K = 6,144$), the Jacobian contains 3.8 billion
elements.  With 99.95\% sparsity, we only store the 1.9 million non-zero
elements---a reduction of three orders of magnitude.
Similarly, the Newton-Kantorovich linear solve requires
$O(K^3 \Nplayers^3)$ operations with dense methods but only
$O(K \Nplayers)$ with sparse iterative solvers like GMRES, exploiting
the fact that matrix-vector products preserve sparsity.

In addition to improving convergence, the analytical Jacobian enables
efficient computation of value function derivatives $\frac{\partial V}{\partial \theta}$
via implicit differentiation:
\begin{equation}
\label{eq:vf_implicit_diff}
\frac{\partial V}{\partial \theta} = \left[I - \frac{\partial T}{\partial V}(V, \theta)\right]^{-1} \frac{\partial T}{\partial \theta}(V, \theta)
\end{equation}
Since the same system matrix appears in both the Newton-Kantorovich step and
implicit differentiation, we factor it once and
reuse this factorization to solve $\dim(\theta)$ linear systems---far more efficient
than numerical differentiation, which would require solving $\dim(\theta)$
additional equilibrium problems.
These derivatives propagate through the choice probabilities and
intensity matrix to provide the log-likelihood gradient for maximum
likelihood estimation, as we demonstrate below in
Section~\ref{sec:loglik}.

\subsubsection{Polyalgorithm for Equilibrium Computation}
\label{sec:vf:newton}

Following \cite{rust87optimal}, we now combine value iteration and
Newton-Kantorovich iterations into a polyalgorithm that leverages the
strengths of each method.
Value iteration establishes a basin of attraction around a stable equilibrium,
then Newton-Kantorovich rapidly converges to the solution.
The algorithm monitors the convergence rate $r^{(n)}$ during value iteration
and switches to Newton-Kantorovich when $r^{(n)}$ approaches the target
modulus $\beta$.

\begin{algorithm}[tbh]
\caption{Polyalgorithm with Convergence Monitoring}
\label{alg:poly}
\begin{algorithmic}[1]
\State Initialize $V^{(0)} = 0$ (or use a previous solution)
\For{$n = 0, 1, 2, \ldots$ and $n < $ \texttt{MAX\_VF\_ITER}}\Comment{\textit{Phase 1: Value Iteration}}
  \State $V^{(n+1)} = T(V^{(n)})$ \Comment{Equilibrium operator}
  \State Compute $\text{diff}^{(n)} = \|V^{(n+1)} - V^{(n)}\|_\infty$
  \If{$\text{diff}^{(n)} < \epsilon_{V}$} \textbf{break} \Comment{Converged}
  \EndIf
  \If{$n \geq $ \texttt{MIN\_MONITORING\_ITER}} \Comment{Convergence monitoring}
    \State Compute $r^{(n)} = \text{diff}^{(n)} / \text{diff}^{(n-1)}$ \Comment{Convergence rate}
    \If{$r^{(n)} > \beta - \epsilon_r$} \textbf{break} \Comment{NFXP switching}
    \EndIf
  \EndIf
\EndFor
\State Set $n_2 \gets 0$ \Comment{\textit{Phase 2: Newton-Kantorovich}}
\While{$\|V^{(n)} - T(V^{(n)})\|_\infty > \epsilon_{V}$ and $n_2 < $ \texttt{MAX\_NEWTON\_ITER}}
  \State $\text{residual} \gets V^{(n)} - T(V^{(n)})$
  \State $J \gets I - \frac{\partial T}{\partial V}(V^{(n)})$ \Comment{Sparse Jacobian}
  \State Solve $J \cdot \Delta V = \text{residual}$ for $\Delta V$ \Comment{Sparse linear solver}
  \State $V^{(n+1)} = V^{(n)} - \Delta V$ \Comment{N-K update}
  \State Set $n_2 \gets n_2 + 1$ and $n \gets n + 1$
\EndWhile
\end{algorithmic}
\end{algorithm}

\begin{figure}[tbhp]
\centering
\begin{subfigure}{0.49\textwidth}
\centering
\includegraphics[width=\textwidth]{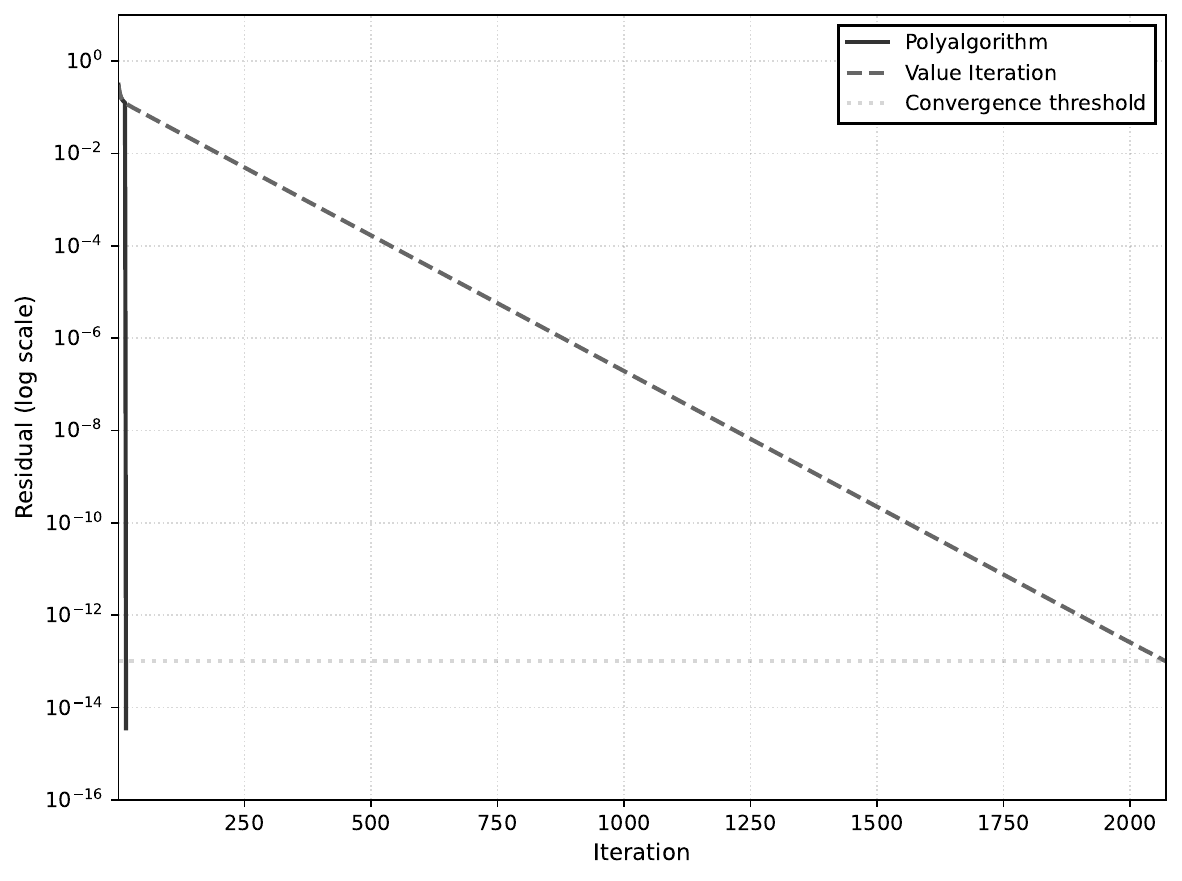}
\caption{Full convergence comparison}
\label{fig:vf_algorithms:full}
\end{subfigure}
\hfill
\begin{subfigure}{0.49\textwidth}
\centering
\includegraphics[width=\textwidth]{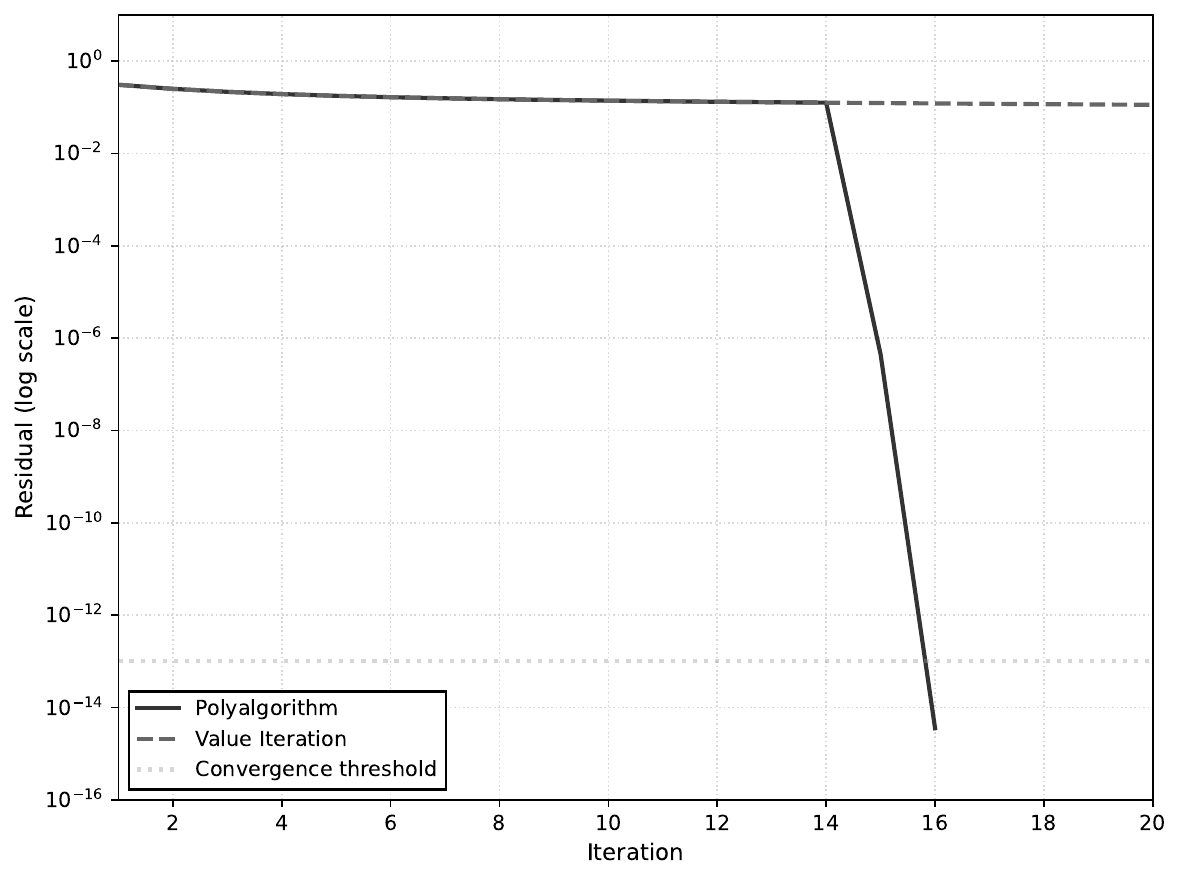}
\caption{First 20 iterations}
\label{fig:vf_algorithms:zoom}
\end{subfigure}
\caption{Value Function Algorithm Convergence Comparison.
Panel (a) shows the complete convergence path for both algorithms,
while panel (b) focuses on the initial iterations to highlight
the early convergence behavior.
The polyalgorithm combines value iteration with Newton-Kantorovich
methods, switching when the convergence rate approaches the
target contraction modulus $\beta$.
The residual is defined as $\|V^{(n)} - T(V^{(n)})\|_{\infty}$, measuring
the maximum absolute difference in the fixed point condition.}
\label{fig:vf_algorithms}
\end{figure}

\begin{figure}[tbhp]
\centering
\includegraphics[width=0.65\textwidth]{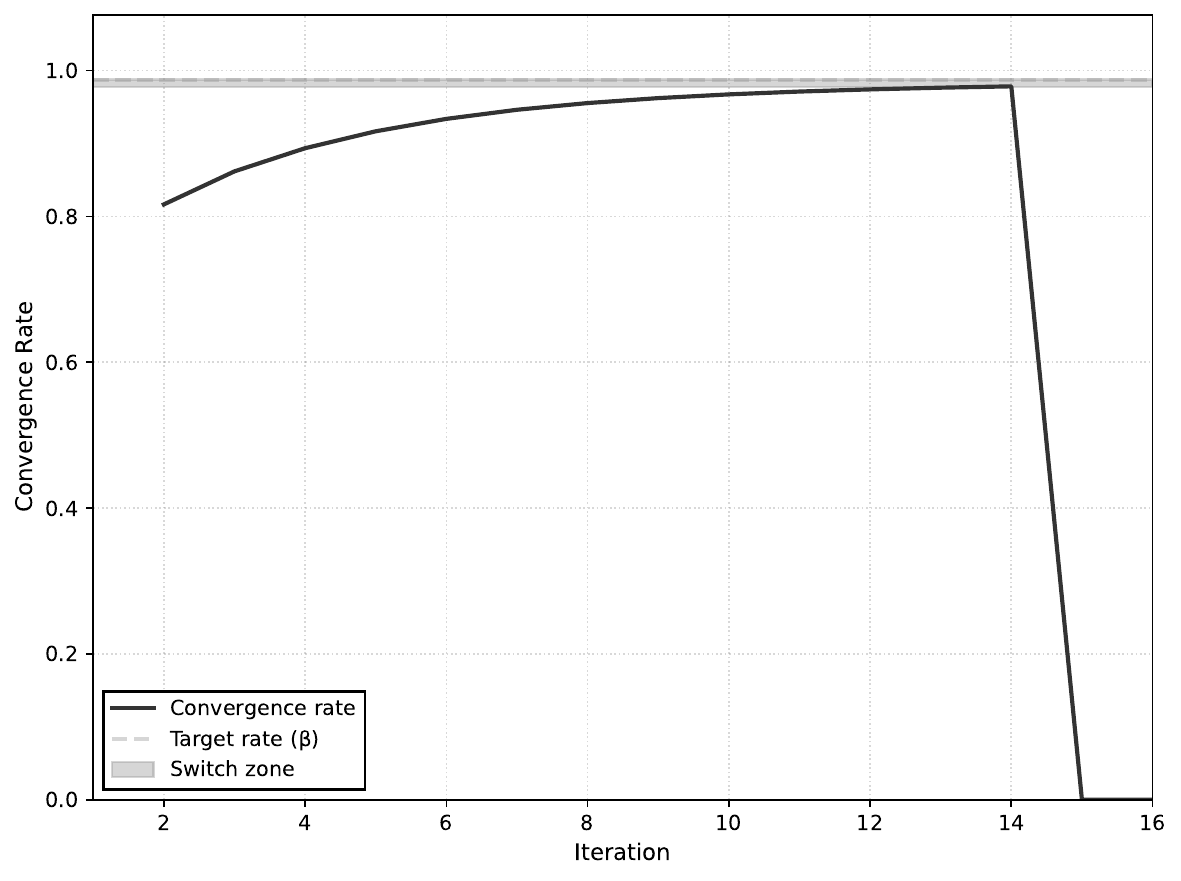}
\caption{Rate of Convergence and Polyalgorithm Switching Criterion.
The figure shows the convergence rate
$r^{(n)} = \|V^{(n)} - V^{(n-1)}\|_{\infty} / \|V^{(n-1)} - V^{(n-2)}\|_{\infty}$
during value iteration, along with the theoretical target rate $\beta$.
The shaded region indicates the switching condition
$r^{(n)} > \beta - \epsilon_r$ (with $\epsilon_r = 0.01$ in this example),
triggering the transition to the Newton-Kantorovich phase.}
\label{fig:vf_rate_analysis}
\end{figure}

Algorithm~\ref{alg:poly} presents the complete polyalgorithm,
which monitors the convergence rate and switches from value iteration to
Newton-Kantorovich when the target rate is approached.
The algorithm requires specifying the convergence tolerance $\epsilon_V$ for the value function,
rate tolerance $\epsilon_r$ for switching, and a minimum number of iterations
\texttt{MIN\_MONITORING\_ITER} before monitoring begins.
For example, in our Monte Carlo experiments we use $\epsilon_V = 10^{-13}$.
We apply value iteration for at least 10 iterations before convergence
monitoring, and we employ a relatively aggressive switching tolerance
$\epsilon_r = 0.1$.

Figure~\ref{fig:vf_algorithms} demonstrates the polyalgorithm's convergence
compared to pure value iteration.
It achieves convergence in significantly
fewer iterations (16 vs 2,071) by leveraging quadratic convergence of Newton-Kantorovich
once the contraction rate approaches the theoretical modulus.
Figure~\ref{fig:vf_rate_analysis} shows the convergence rate relative to $\beta$,
with the switching condition met at iteration 14.

\subsection{Uniform Representation of the Value Function}
\label{sec:vf:uniform}

We now develop a novel uniform representation of the value function
based on the uniformization method from Section~\ref{sec:model:ctmc}.
While the representation builds on the CCP inversion result of
\cite{hotz93conditional} as applied to continuous-time settings by
\citet*{abbe-2016}, our uniformization-based approach reveals a clear
connection to discrete-time models and enables us to establish
convergence rates for policy evaluation.

\begin{theorem}[Uniform Policy Evaluation]
  \label{thm:pe:uniform}
  Suppose \allassumptions{} hold.
  Given a profile of choice probabilities $\sigma$,
  $V_i$ satisfies the fixed-point equation
  defined by the \emph{uniform policy evaluation operator}:
  \begin{equation}
    \label{eq:bellman_uniform}
    V_i = \overline{\Gamma}_{i}^{\sigma} V_i \equiv U_i(\sigma) + \overline\beta_i \Sigma(\sigma) V_i,
  \end{equation}
  where
  \begin{equation*}
    U_i(\sigma) \equiv \frac{1}{\rho_i + \unifrate} \left[ u_i + L_i C_i(\sigma_i) \right],\quad
    \overline\beta_i \equiv \frac{\unifrate}{\rho_i + \unifrate} \in (0,1),\quad
    \Sigma(\sigma) \equiv I + \frac{Q(\sigma)}{\unifrate},
  \end{equation*}
  $L_i = \diag(\lambda_{i1}, \dots, \lambda_{iK})$ is the diagonal matrix
  of decision arrival rates for player $i$ in each state,
  $C_i(\sigma_i)$ is the $K \times 1$ vector with $k$-th element
  $\sum_{j=0}^{\Nchoices-1} \sigma_{ijk} \left[ \psi_{ijk} + e_{ijk}(\sigma_i) \right]$,
  where $e_{ijk}(\sigma_i) \equiv \E[\varepsilon_{ijk} \mid j \in \argmax_{j'} \{\psi_{ij'k} + \varepsilon_{ij'k} + V_{i,l(i,j',k)}\}]$
  is the conditional expectation of the choice-specific shock given that action $j$ is chosen
  (reflecting selection effects from optimal choice),
  and $Q(\sigma)$ is the intensity matrix given $\sigma$.
  The operator $\overline{\Gamma}_{i}^{\sigma}$ is a contraction mapping with respect
  to the supremum norm with modulus $\overline\beta_i$.
\end{theorem}

Theorem~\ref{thm:pe:uniform} reveals a direct connection to discrete-time
dynamic discrete choice models through uniformization.
The representation defines a contraction with modulus $\overline\beta_i$,
the effective discount factor, which depends on the uniformization rate
$\unifrate$ relative to the discount rate $\rho_i$.
Under Assumptions~\ref{assp:rho} and \ref{assp:rates}, $\overline\beta_i \in (0,1)$.
The effective flow utility $U_i(\sigma)$ combines the actual flow
utility with the rate-weighted expected instantaneous payoff at the
agent's next decision time, while $\Sigma(\sigma)$ is a stochastic
matrix serving as the effective transition probability matrix.

\begin{remark}[Relationship to Existing Representations]
Our uniform operator $\overline{\Gamma}_i^{\sigma}$ relates to existing representations in two ways.
First, it can be viewed as the uniformized version of the recursive policy evaluation
operator $\Gamma_i^{\sigma}$ from \citet[][eq.~8]{ctgames}:
\begin{equation*}
\Gamma_i^{\sigma}(V_i) = D_i \left[ u_i + \tilde{Q}_0 V_i + \sum_{m \neq i} L_m \Sigma_m(\sigma_m) V_i
  + L_i \left\{ \Sigma_i(\sigma_i) V_i + C_i(\sigma_i) \right\}\right],
\end{equation*}
where $D_i$ is a diagonal matrix with elements $(D_i)_{kk} = 1/(\rho_i + \unifrate_k)$,
$\tilde{Q}_0 = Q_0 - \diag(q_{11}, \dots, q_{KK})$ is the off-diagonal part of $Q_0$,
and $\Sigma_{m}(\sigma_{m})$ is the $K \times K$ transition matrix with $(k,k')$ element
equal to the probability of transitioning from state $k$ to state $k'$ due to
player $m$'s action given CCPs $\sigma_{m}$ and continuation
state function $l(m,\cdot,\cdot)$.

Second, \citet*{abbe-2016} and \cite{ctgames} derived non-recursive closed-form
representations, with the latter generalized to allow heterogeneous rates:
\begin{equation*}
V_i = \left[\rho_i I + \sum_{m=1}^{\Nplayers} L_m [I - \Sigma_m(\sigma_m)] - Q_0\right]^{-1} [u_i + L_i C_i(\sigma_i)].
\end{equation*}
This representation solves for $V_i$ explicitly and is useful for
identification and two-step estimation.
In contrast, our recursive representation in \eqref{eq:bellman_uniform}
directly connects to discrete-time models through
the probability transition matrix $\Sigma(\sigma)$
and contraction modulus $\overline{\beta}_i < 1$.
\end{remark}

\begin{remark}[Computation of $C_i(\sigma_i)$]
\citet*{abbe-2016} showed that the conditional expectation
$e_{ijk}(\sigma_i)$ appearing in $C_i(\sigma_i)$
can be expressed purely as a function of player $i$'s own choice
probabilities $\sigma_i$, as opposed to the full profile $\sigma$.
Under common distributional assumptions (type 1 extreme value and binary normal),
$e_{ijk}(\sigma_i)$ has closed-form expressions \citep[][Lemma 1]{ctgames}.
\end{remark}

\begin{lemma}[Uniform and Non-Uniform Contraction Moduli]
\label{lem:contraction_moduli}
The contraction modulus $\beta_i$ from Theorem~\ref{thm:vi:non-uniform} and
the uniform policy evaluation modulus $\overline{\beta}_i$ from
Theorem~\ref{thm:pe:uniform} satisfy $\beta_i \leq \overline{\beta}_i$,
with equality when $\unifrate_k$ is constant across all states $k$
and $\unifrate = \max_k \unifrate_k$.
\end{lemma}

The lemma establishes that (non-uniform) value iteration converges at least as
fast as uniform policy evaluation.  The uniform representation trades some
tightness in the contraction bound for the computational advantages of working
with a stochastic matrix $\Sigma$ and the connection to discrete-time
methods.

This section has developed computational methods for \emph{solving}
continuous-time dynamic discrete choice games,
including a polyalgorithm for computing equilibria and a
uniform policy evaluation representation that connects to
discrete-time models.
With these established, we now turn to the computational aspects
of \emph{estimating} such games using only discrete-time data.

\section{Computing the Log-Likelihood Function and its First Derivatives}
\label{sec:loglik}

\subsection{Identification with Snapshot Data}

While this paper focuses on computational methods for estimation,
\citet{ctgames} provides a comprehensive treatment of
identification in this framework.
The identification analysis proceeds in two steps: first recovering the
intensity matrix $Q$ from discrete-time transition probabilities $P(\Delta)$,
then recovering structural primitives (payoffs, value functions, and
heterogeneous arrival rates $\lambda_{ik}$) from $Q$.

The first step addresses the aliasing problem: multiple intensity matrices
$Q$ may generate the same discrete-time transitions $P(\Delta) = \exp(\Delta Q)$.
However, the structural model provides natural restrictions that ensure
identification.
For instance, in entry/exit models, firms cannot force rivals to enter or exit,
implying zeros in specific positions of the $Q$ matrix.
Similarly, exogenous state variables (like demand conditions) can only be
changed by nature, not by players' actions.
These exclusion restrictions, arising naturally from the economic model,
restrict the domain of possible $Q$ matrices to ensure unique identification
from $P(\Delta)$.
Alternative approaches include using sufficiently small sampling intervals $\Delta$
where $P(\Delta) \approx I + \Delta Q$, or observing transitions at multiple
distinct intervals.

Given $Q$, the structural primitives can be identified with $2K$ linear
restrictions per player---substantially fewer than the exponentially many
restrictions required in discrete-time models \citep{pesendorfer08asymptotic}.
This reduction occurs because continuous-time models avoid simultaneous moves:
only one player acts at any instant, simplifying the identification problem.
Common restrictions include normalizations (e.g., $\psi_{i0k} = 0$ for continuation),
constant parameters across states (e.g., entry costs independent of demand),
or known values in terminal states.
Importantly, heterogeneous arrival rates $\lambda_{ik}$ are not
automatically identified from discrete-time data and require these
additional restrictions.

The sampling frequency $\Delta$ affects both identification and estimation.
Smaller $\Delta$ aids identification by making $P(\Delta)$ closer to the
underlying $Q$, but the computational methods we discuss next are designed
to handle the matrix exponential calculations for any fixed $\Delta$.

\subsection{Likelihood Computation with Discrete-Time Data}

Estimation of the model with continuous-time data is relatively
straightforward, so here we focus on the more computationally
difficult task of estimation with discrete-time data.\footnote{With
  continuous-time data where exact transition times are observed, estimation
  is computationally simpler as the likelihood depends directly on the intensity
  matrix $Q$ rather than requiring matrix exponential calculations
  \citep{billingsley61statistical}.
  Our focus on discrete-time snapshot data addresses the computationally more
  difficult case that is common in empirical practice.}
This is the typical case in many empirical applications where data are collected
annually, quarterly, or monthly, but the underlying process evolves continuously.
Suppose the researcher observes snapshots of markets at
regularly-spaced intervals $\Delta > 0$.
Let $M$ denote the number of markets and $T_m$ the number of snapshots
over time observed in each market $m = 1, \dots, M$.
The sample can be represented as $\lbrace k_{m,0}, \dots, k_{m,T_m} \rbrace_{m = 1}^{M}$,
where $k_{mn}$ denotes the observed state for snapshot $n$ in market $m$.
Given the sample, we first pre-calculate
transition counts $d_{kk'}$ for all pairs of states $(k, k')$:
\begin{equation*}
  d_{kk'} \equiv \sum_{m=1}^M \sum_{n=1}^{T_m} \1\left\lbrace k_{m,n-1} = k, k_{m,n} = k' \right\rbrace.
\end{equation*}
The log-likelihood function can then be expressed simply in terms of these counts as
\begin{equation}
  \ell(\theta)
  = \ln \prod_{m=1}^M \prod_{n=1}^{T_m} P\left(\Delta, \theta\right)_{k_{m,n-1}, k_{m,n}}
  = \sum_{k'=1}^K d_{k'}^\top \ln P\left(\Delta, \theta\right) e_{k'} \label{eq:loglik:dt}
\end{equation}
where $d_{k'}^\top = ( d_{1k'}, \dots, d_{Kk'} )$ contains transition counts
into state $k'$ and $e_k$ is the $k$-th basis vector.

The form of \eqref{eq:loglik:dt} enables several important computational efficiencies.
First, we need to compute only those columns of $P(\Delta,\theta)$ corresponding to destination states
$k'$ with positive observed transition counts.
Individual columns can be obtained through matrix-vector products of the form
$P(\Delta,\theta) e_{k'}$, substantially reducing the
computational burden when the number of observed destination states is much smaller
than the size of the state space---a common scenario in large models with moderate sample
sizes.
Second, for each required column $k'$, we only need the inner product
$d_{k'}^\top \ln P(\Delta,\theta) e_{k'}$ using the non-zero elements of $d_{k'}$.

The gradient of $\ell(\theta)$ will also be important for accurate
and efficient optimization:
\begin{equation}
 \label{eq:loglik:dt:grad}
  \frac{\partial \ell}{\partial \theta}(\theta)
  = \sum_{k=1}^K \sum_{k'=1}^K \frac{d_{kk'}}{P\left(\Delta, \theta \right)_{kk'}} \frac{\partial P\left(\Delta, \theta \right)_{kk'}}{\partial \theta}
\end{equation}

Under standard regularity conditions and assuming a unique equilibrium,
the maximum likelihood estimator obtained by maximizing \eqref{eq:loglik:dt}
has the usual asymptotic properties: consistency, asymptotic normality,
and efficiency \citep{newey94large}.
Our analytical derivatives provide the exact score function
$\nabla \ell(\theta)$ in \eqref{eq:loglik:dt:grad}, eliminating numerical
approximation error in both optimization and standard error computation.

Recall that the transition matrix $P(\Delta, \theta)$ appearing
in \eqref{eq:loglik:dt} and \eqref{eq:loglik:dt:grad}
is computed via the matrix exponential as $P(\Delta,\theta) = \exp(\Delta Q(\theta))$.
Therefore, to maximize the log-likelihood function we shall require an
efficient method to compute $\exp(\Delta Q(\theta)) v$,
the action of the matrix exponential on a probability vector $v$,
along with the derivatives
$\frac{\partial}{\partial \theta}\exp(\Delta Q(\theta))v$.
The following section presents a numerically stable uniformization-based algorithm
to compute these quantities efficiently.

Given a valid uniformization rate $\unifrate$ satisfying \eqref{eq:unifrate},
define $\Sigma \equiv I + Q / \unifrate$,
so that $\Delta Q = \unifrate \Delta \Sigma - \unifrate \Delta I$.
Then following \eqref{eq:expm:unif}, the action of $\exp(\Delta Q)$
on a vector $v$ can be written as
\begin{equation}
  \label{eq:expmv}
  \exp(\Delta Q)v = \exp(\unifrate \Delta \Sigma - \unifrate \Delta I)v = \e^{-\unifrate \Delta} \sum_{j=0}^\infty \frac{(\unifrate \Delta)^j \Sigma^j v}{j!}.
\end{equation}
Importantly, all elements of $\Sigma$ are non-negative, meaning that this
calculation will not suffer from cancellation of alternating positive and
negative terms \citep{goldberg-1991}.
As a result, computations will be much more numerically stable for the
uniformization of a rate matrix $Q$ than for the exponential of a
generic matrix $A$.
Furthermore, we can compute \eqref{eq:expmv} efficiently using the
following recurrence:
\begin{equation*}
  \exp(\Delta Q) v
  = \e^{-\unifrate \Delta} \sum_{j=0}^\infty \accumv_j,
  \quad \accumv_0 = v, \quad\text{and}\quad
  \accumv_j = \frac{\unifrate \Delta \Sigma}{j} \accumv_{j-1},
\end{equation*}
with $\accumv_j$ involving only a single sparse matrix vector product
\citep{sherlock-2022}.

In practice, we truncate the series at $j = \bar{J}_\varepsilon < \infty$ such that
the approximation error is below some tolerance $\varepsilon$.
Since uniformization produces a Poisson process, we determine $\bar{J}_{\varepsilon}$
using Poisson tail probabilities \citep{fox-glenn-1988, reibman-trivedi-1988}.
When $\Sigma$ is a stochastic matrix and $v$ is a probability vector, each term
$\Sigma^j v$ is also a probability vector with $\norm{\Sigma^j v}_1 = 1$.
This property allows us to bound the truncation error directly: we choose
$\bar{J}_\varepsilon$ as the smallest integer such that
\begin{equation*}
\e^{-\unifrate \Delta} \sum_{j=\bar{J}_\varepsilon + 1}^{\infty} \frac{(\unifrate \Delta)^j}{j!} = 1 - \text{PoissonCDF}(\bar{J}_{\varepsilon}; \unifrate \Delta) <
\varepsilon,
\end{equation*}
which gives $\bar{J}_{\varepsilon} = \text{InvPoissonCDF}(1-\varepsilon; \unifrate \Delta)$.\footnote{\cite{sherlock-2022} discusses accurate computation of Poisson tail
probabilities for small $\varepsilon$ and modifications to handle overflow in
$\sum_{j=0}^J \frac{(\unifrate \Delta)^j}{j!}$.}

Uniformization of the matrix exponential also leads to an efficient,
recursive algorithm for computing its derivatives \citep{rupp-2024}.
Returning to \eqref{eq:expmv} and differentiating, noting that
$\frac{\partial (\Sigma^0)}{\partial \theta} = \frac{\partial I}{\partial \theta} = 0$,
so we can start the sum at $j=1$:
\begin{equation}
  \label{eq:expm:deriv:vec}
  \frac{\partial \exp(\Delta Q)}{\partial\theta} v
  = \e^{-\unifrate \Delta} \sum_{j=1}^\infty \frac{(\unifrate \Delta)^j}{j!} \frac{\partial \Sigma^j}{\partial \theta} v.
\end{equation}
Note that $\frac{\partial \Sigma^j}{\partial \theta}$ denotes the derivative of the
matrix $\Sigma^j$, as opposed to the $j$-th power of the derivative of $\Sigma$.
Computing this directly is infeasible,
but we can compute it recursively from $\frac{\partial \Sigma}{\partial \theta}$
by noting that $\Sigma^j = \Sigma \Sigma^{j-1}$ and applying
the product rule:
\begin{equation}
\label{eq:expm:deriv:rec}
\frac{\partial \Sigma^j}{\partial \theta}
= \frac{\partial \Sigma}{\partial \theta} \Sigma^{j-1} + \Sigma \frac{\partial \Sigma^{j-1}}{\partial \theta}.
\end{equation}
Substituting \eqref{eq:expm:deriv:rec} into \eqref{eq:expm:deriv:vec}:
\begin{align*}
  \frac{\partial \exp(\Delta Q)}{\partial\theta} v
  &= \e^{-\unifrate \Delta} \sum_{j=1}^\infty \left[
    \frac{(\unifrate \Delta)^j}{j!} \frac{\partial \Sigma}{\partial \theta} \Sigma^{j-1} v
    + \frac{(\unifrate \Delta)^j}{j!} \Sigma \frac{\partial \Sigma^{j-1}}{\partial \theta} v
  \right] \\
  &= \e^{-\unifrate \Delta} \sum_{j=1}^\infty \left[
    \frac{\unifrate \Delta}{j} \frac{\partial \Sigma}{\partial \theta} \times \underbrace{\frac{(\unifrate \Delta \Sigma)^{j-1}}{(j-1)!} v}_{= \accumv_{j-1}}
    + \frac{\unifrate \Delta \Sigma}{j} \times \underbrace{\frac{(\unifrate \Delta)^{j-1}}{(j-1)!} \frac{\partial \Sigma^{j-1}}{\partial \theta} v}_{\equiv \delta_{j-1}}
  \right] \\
  &= \e^{-\unifrate \Delta} \sum_{j=1}^\infty \delta_{j},
\end{align*}
where we can compute $\accumv_{j}$ and $\delta_{j}$ recursively as follows:
\begin{align*}
  \accumv_{0} &= v,\\
  \delta_{0} &= 0,\\
  \accumv_{j} &= \frac{\unifrate \Delta \Sigma}{j} \accumv_{j-1}, \\
  \delta_{j} &= \frac{\unifrate \Delta}{j} \left[ \frac{\partial \Sigma}{\partial \theta} \accumv_{j-1} + \Sigma \delta_{j-1} \right].
\end{align*}

A pseudocode implementation is presented below as Algorithm~\ref{alg:expmvd}.

\begin{remark}
In recursions involving sparse matrix products, the resulting matrices can be subject
to fill-in, increasing the storage and floating point operations required substantially.
Importantly, for both $\accumv_{j}$ and $\delta_{j}$ we only need to store $K \times 1$ vectors.
The only matrices stored are the sparse matrices $\Sigma$ and
$\frac{\partial \Sigma}{\partial \theta}$.
Each iteration of the recursion only involves three sparse matrix-vector
products.
\end{remark}

\begin{remark}
  The term $\accumv_{j}$ is also used for computing $\exp(\Delta Q) v$,
  so the matrix exponential times $v$ and its derivatives can be computed simultaneously.
  When $\theta$ contains multiple parameters, we can compute the derivatives simultaneously
  in the same loop simply by storing separate vectors $\delta_{j,\alpha}$
  for each component $\alpha$ of $\theta$, since $\accumv_{j}$ is independent of $\theta$.
\end{remark}

\begin{algorithm}[htbp]
  \caption{Uniformization Algorithm for Matrix Exponential and Derivatives}
  \label{alg:expmvd}
  \begin{algorithmic}[1]
  \Statex \textbf{Input:} $Q$ (intensity matrix), $\frac{\partial Q}{\partial \theta}$ (derivative), $\Delta$ (interval), $v$ (vector), $\varepsilon$ (tolerance)
  \Statex \textbf{Output:} $\exp(\Delta Q) v$ and $\frac{\partial \exp(\Delta Q)}{\partial \theta} v$
  \Function{expmvd}{$Q, \frac{\partial Q}{\partial \theta}, \Delta, v, \varepsilon$}
  \State $\unifrate \gets \max(\text{abs}(\text{diag}(Q)))$  \Comment{Uniformization rate}
  \State $S \gets \Delta Q + \unifrate \Delta I$  \Comment{Scaled transition matrix: $S = \unifrate \Delta \Sigma$}
  \State $D \gets \Delta \frac{\partial Q}{\partial \theta}$  \Comment{Scaled derivative: $D = \unifrate \Delta \frac{\partial \Sigma}{\partial \theta}$}
  \State $\bar{J}_{\varepsilon} \gets \text{InvPoissonCDF}(1 - \varepsilon, \unifrate \Delta)$
  \State $\accumv \gets v$
  \State $\text{expQv} \gets \accumv$
  \State $\delta \gets 0_{K \times 1}$
  \State $\text{dexpQv} \gets 0_{K \times 1}$
  \For{$j \gets 1$ \textbf{to} $\bar{J}_{\varepsilon}$}
      \State $\delta \gets (D\, \accumv + S\, \delta) / j$  \Comment{Compute $\delta_j$ using $\accumv_{j-1}$}
      \State $\text{dexpQv} \gets \text{dexpQv} + \delta$
      \State $\accumv \gets S\, \accumv / j$  \Comment{Update to $\accumv_j$}
      \State $\text{expQv} \gets \text{expQv} + \accumv$
  \EndFor
\State $\text{expQv} \gets \exp(-\unifrate \Delta) \times \text{expQv}$
\State $\text{dexpQv} \gets \exp(-\unifrate \Delta) \times \text{dexpQv}$
\State \Return $\text{expQv}$, $\text{dexpQv}$
\EndFunction
\end{algorithmic}
\end{algorithm}

Finally, we return to computing the derivatives
$\frac{\partial Q}{\partial \theta}$
which are required as inputs for the recursive algorithm above.
These involve both direct parameter effects and indirect effects
through choice probabilities:
\begin{equation}
\label{eq:chainrule:Q}
\frac{\partial Q}{\partial \theta} = \frac{\partial Q}{\partial \sigma}\frac{\partial \sigma}{\partial \theta} + \left.\frac{\partial Q}{\partial \theta}\right|_{\sigma},
\end{equation}
where the subscript notation indicates the direct effect holding $\sigma$ fixed.
Each element of $Q$ has the form:
\begin{equation*}
q_{kk'} = \begin{cases}
q_{0kk'} + \sum_{i} \sum_{j: l(i,j,k) = k'} \lambda_{ik} \sigma_{ijk} & \text{if } k \neq k' \\
-\sum_{k' \neq k} q_{kk'} & \text{if } k = k'
\end{cases}
\end{equation*}
with diagonal elements determined by the row-sum constraint.
For off-diagonal elements ($k \neq k'$), the total derivatives combining both direct and indirect effects are
\begin{equation}
\label{eq:loglik:intensity_deriv}
\frac{\partial q_{kk'}}{\partial \theta} = \frac{\partial q_{0kk'}}{\partial \theta} + \sum_{i} \sum_{j: l(i,j,k) = k'}
\left( \frac{\partial \lambda_{ik}}{\partial \theta} \sigma_{ijk} + \lambda_{ik} \frac{\partial \sigma_{ijk}}{\partial \theta} \right),
\end{equation}
where the first two terms ($\frac{\partial q_{0kk'}}{\partial \theta}$ and $\frac{\partial \lambda_{ik}}{\partial \theta} \sigma_{ijk}$)
capture direct parameter dependence of transition rates (e.g., $\gamma$ or $\lambda_{ik}$),
and the final term ($\lambda_{ik} \frac{\partial \sigma_{ijk}}{\partial \theta}$) captures the
indirect effect through equilibrium choice probability derivatives, derived below.
Diagonal element derivatives follow from the row-sum constraint:
$\frac{\partial q_{kk}}{\partial \theta} = -\sum_{k' \neq k} \frac{\partial q_{kk'}}{\partial \theta}$.

Recall that the equilibrium choice probabilities are determined by the
Markov perfect equilibrium condition.
For type I extreme value errors,
these take the familiar logit form:
\begin{equation*}
\sigma_{ijk} = \frac{\exp(V_{i,l(i,j,k)} + \psi_{ijk})}{\sum_{j'=0}^{\Nchoices-1} \exp(V_{i,l(i,j',k)} + \psi_{ij'k})}.
\end{equation*}
The logit structure provides closed-form derivatives.
Differentiating with respect to continuation values yields:
\begin{equation*}
  \frac{\partial \sigma_{ijk}}{\partial V_{i,l(i,j,k)}} = \sigma_{ijk}(1-\sigma_{ijk})
  \quad\text{and}\quad
  \frac{\partial \sigma_{ijk}}{\partial V_{i,l(i,j',k)}} = -\sigma_{ijk}\sigma_{ij'k}
  \text{ for } j' \neq j.
\end{equation*}
That is, the derivative with respect to action $j$'s own continuation value is
$\sigma_{ijk}(1-\sigma_{ijk})$, while the derivative with respect to a different action's
continuation value is $-\sigma_{ijk}\sigma_{ij'k}$.
The derivatives $\frac{\partial \sigma}{\partial \psi}$ follow the same functional form,
while $\frac{\partial \psi}{\partial \theta}$ depends on the specific parameter
(e.g., entry costs).

This completes the computational pathway for evaluating both
the log-likelihood function and its gradient with respect to the structural parameters,
allowing exact gradient computation for maximum likelihood estimation
while exploiting sparsity and uniformization for efficiency and numerical stability.

\section{Monte Carlo Evidence on Finite-Sample Performance}
\label{sec:mc}

We evaluate the finite-sample properties of our estimation methods through
Monte Carlo simulations, focusing on how analytical derivatives improve
both statistical accuracy and computational efficiency.
As is common in empirical practice, we work with discrete-time snapshot
data where the econometrician observes states only at fixed intervals $\Delta$,
making matrix exponential computation necessary for likelihood evaluation.

\subsection{Model Specification and Estimation}

The model for our Monte Carlo experiments is based on the example model of
entry and exit with $\Nplayers$ firms operating in a single product market.
Firms operate under stochastically varying market demand conditions with
$\Ndemand$ demand levels $x_{k0} \in \{ 0, \dots, \Ndemand - 1 \}$.
The payoff relevant states are the number of active firms,
denoted $\nactivek$,
and the current state of market demand, $x_{k0}$.\footnote{In principle
  we could use this to reduce the state space of the model along the
  lines of \citet*{abbe-2016}, but we kept the full state space to
  keep the source code easy to understand and better illustrate how
  the computational complexity varies with the number of firms.}
For our experiments, we specify $\Nplayers = 7$ players and
$\Ndemand = 5$ demand states, yielding $K = 2^{\Nplayers} \times \Ndemand = 640$
total states.

Three structural parameters determine the flow payoffs and
instantaneous payoffs:
$\thetaEC$ is the entry cost (incurred when inactive firms enter),
$\thetaRN$ represents competitive effect, and
$\thetaD$ represents the profitability of the market level of demand,
$x_{k0}$.
For firm $i$ in state $k$, the flow payoff $u_{ik}$ is
\begin{equation*}
  u_{ik} = x_{ki} \times \left(\thetaRN \nactivek + \thetaD x_{k0} \right),
\end{equation*}
where $x_{ki} \in \{0,1\}$ is the activity indicator for firm $i$
when the market state is $k$.
The exit scrap value is normalized to zero.

We generate discrete-time datasets with sampling interval $\Delta=1.0$ and
two sample sizes: $T \in\{ 1000, 4000 \}$ observations per replication.
The true parameter values are given in Table~\ref{tab:mc:7x5:1000}.
With $\lambda = 1.0$, each player makes on average one decision per
observation interval.\footnote{Both numerical and analytical gradient
  approaches use the same starting values, which are chosen to be far
  from the true parameters:
  $(\thetaEC, \thetaRN, \thetaD, \lambda, \gamma) = (-1.0, -0.1, 1.0, 0.2, 1.0)$
  compared to true values of $(-2.0, -0.5, 2.0, 1.0, 0.3)$.
  The value function convergence tolerance is set to $10^{-13}$ for all experiments.}

As noted in Section~\ref{sec:vf}, the equilibrium system operator $T(V)$
may not be globally contractive, allowing for multiple equilibria.
However, for this specification there appears to be a unique equilibrium.\footnote{
  We searched for multiple equilibria by solving the equilibrium system $T(V)$
  from 10,000 random initial value vectors with components uniformly distributed
  over $[\min u_{ik} / \rho, \max u_{ik} / \rho]$ using Newton-Kantorovich
  iterations.  We solved the system to a tolerance of $10^{-13}$, then
  clustered the choice probability vectors using a tolerance of $10^{-3}$.
  Each of the solutions converged to the same equilibrium.
}
This allows us to focus on evaluating our matrix exponential and analytical
derivative methods for a given equilibrium.

We compare three estimation approaches over 100 Monte Carlo replications:
\begin{enumerate}
\item \textbf{Numerical Gradient}: L-BFGS-B with finite-difference gradients.
\item \textbf{Analytical Gradient}: L-BFGS-B with exact gradients from Algorithm~\ref{alg:expmvd}.
\item \textbf{Infeasible Start}: Best result from initialization at true values.
\end{enumerate}
All approaches maximize the log-likelihood function $\ell(\theta)$ in
\eqref{eq:loglik:dt} using L-BFGS-B and sparse matrix routines from
the Python SciPy library.
For the infeasible start results, we initialize the optimization using
the best parameter values from three choices: (1) the true parameter
values, (2) the estimates from the numerical gradient approach, and
(3) the estimates from the analytical gradient approach.

\subsection{Finite-Sample Statistical Properties}

\begin{table}[tbph]
\centering
\caption{Monte Carlo Results (1,000 Observations, 100 Replications)}
\label{tab:mc:7x5:1000}
\begin{tabular}{lrrrrrrr}
\toprule
 &  & \multicolumn{2}{c}{Numerical Gradient} & \multicolumn{2}{c}{Analytical Gradient} & \multicolumn{2}{c}{Infeasible Start} \\
\cmidrule(lr){3-4} \cmidrule(lr){5-6} \cmidrule(lr){7-8}
Parameter & True & Mean & S.D. & Mean & S.D. & Mean & S.D. \\
\midrule
$\thetaEC$ & -2.000 & -1.984 & 0.239 & -1.991 & 0.204 & -2.002 & 0.200 \\
$\thetaRN$ & -0.500 & -0.514 & 0.120 & -0.511 & 0.094 & -0.511 & 0.096 \\
$\thetaD$ & 2.000 & 2.024 & 0.290 & 2.004 & 0.246 & 2.008 & 0.257 \\
$\lambda$ & 1.000 & 1.062 & 0.437 & 1.011 & 0.057 & 1.012 & 0.058 \\
$\gamma$ & 0.300 & 0.301 & 0.018 & 0.301 & 0.017 & 0.301 & 0.017 \\
\midrule
Time (s) &  & 605.7 & 152.3 & 329.4 & 78.7 & 16.0 & 11.5 \\
Iterations &  & 38.4 & 5.9 & 38.6 & 4.9 & 1.4 & 2.4 \\
Func. Eval. &  & 370.9 & 64.4 & 59.8 & 9.6 & 4.1 & 3.9 \\
Log-likelihood &  & -5.1243 & 0.1051 & -5.1185 & 0.1030 & -5.1147 & 0.1046 \\
\bottomrule
\end{tabular}
\end{table}

\begin{table}[tbph]
\centering
\caption{Monte Carlo Results (4,000 Observations, 100 Replications)}
\label{tab:mc:7x5:4000}
\begin{tabular}{lrrrrrrr}
\toprule
 &  & \multicolumn{2}{c}{Numerical Gradient} & \multicolumn{2}{c}{Analytical Gradient} & \multicolumn{2}{c}{Infeasible Start} \\
\cmidrule(lr){3-4} \cmidrule(lr){5-6} \cmidrule(lr){7-8}
Parameter & True & Mean & S.D. & Mean & S.D. & Mean & S.D. \\
\midrule
$\thetaEC$ & -2.000 & -2.005 & 0.093 & -2.004 & 0.093 & -1.998 & 0.091 \\
$\thetaRN$ & -0.500 & -0.500 & 0.046 & -0.500 & 0.046 & -0.502 & 0.046 \\
$\thetaD$ & 2.000 & 1.995 & 0.126 & 1.994 & 0.127 & 1.997 & 0.130 \\
$\lambda$ & 1.000 & 1.001 & 0.028 & 1.001 & 0.028 & 1.002 & 0.027 \\
$\gamma$ & 0.300 & 0.300 & 0.009 & 0.300 & 0.009 & 0.300 & 0.009 \\
\midrule
Time (s) &  & 725.2 & 120.7 & 428.6 & 52.3 & 21.4 & 14.0 \\
Iterations &  & 38.3 & 4.5 & 37.6 & 3.8 & 1.2 & 1.6 \\
Func. Eval. &  & 381.1 & 53.1 & 59.9 & 7.6 & 4.1 & 3.2 \\
Log-likelihood &  & -5.1284 & 0.0533 & -5.1292 & 0.0530 & -5.1286 & 0.0550 \\
\bottomrule
\end{tabular}
\end{table}

\begin{figure}[tbh]
  \begin{center}
    \includegraphics[width=\textwidth]{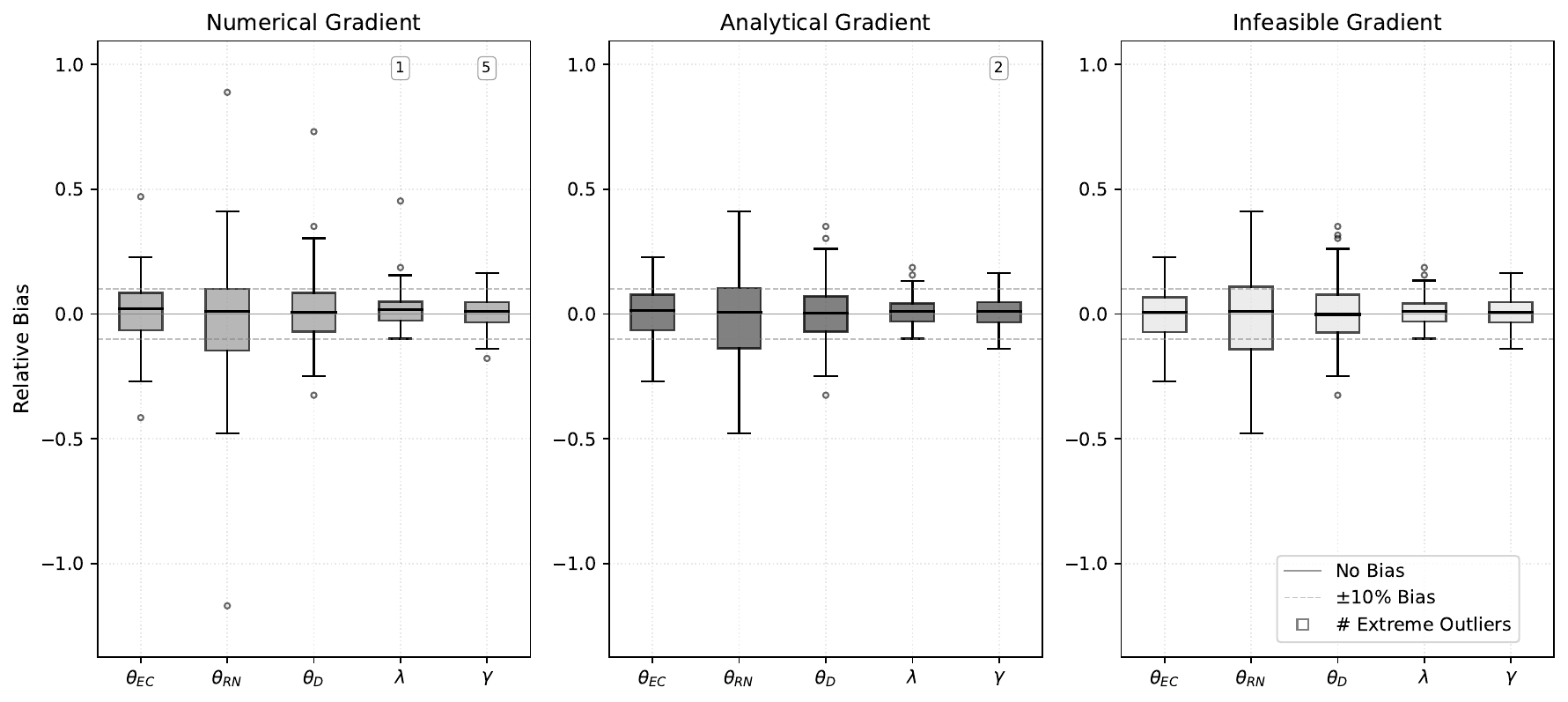}
    \caption{Box plots of relative bias for parameter estimates from
    Monte Carlo simulations (100 replications with 1,000 observations
    per replication).
    Relative bias is defined as $(\hat\theta - \theta_0) / |\theta_0|$.
    Numbers above boxes indicate extreme outliers beyond the plot range,
    which were omitted for better scaling.}
    \label{fig:mc:boxplots:7x5:1000}
  \end{center}
\end{figure}

Tables~\ref{tab:mc:7x5:1000} and \ref{tab:mc:7x5:4000} present results from
100 Monte Carlo replications for sample sizes of 1,000 and 4,000 observations,
respectively.
Our results reveal that beyond computational gains, analytical
derivatives substantially improve small-sample estimation performance.

With the smaller sample size (Table~\ref{tab:mc:7x5:1000}), analytical gradients
show dramatically better performance in terms of variance.
The standard deviation of the $\thetaEC$ estimates is 0.239 with numerical
gradients versus 0.204 with analytical gradients.
More strikingly, the estimates of $\lambda$ show standard deviations of 0.437
versus 0.057, respectively---a nearly eight-fold improvement.
This dramatic difference suggests that numerical gradient approximation errors
contaminate the optimization process, preventing the optimizer from reaching
the true maximum and introducing additional variance into the estimates.
The analytical gradient results are very similar to the infeasible baseline
results, suggesting that the analytical gradient method achieves near-optimal
convergence.

The outliers from using numerical gradients can be seen clearly in the box
plots shown in Figure~\ref{fig:mc:boxplots:7x5:1000}.
With analytical gradients, we have fewer extreme outliers, providing
substantial improvements in finite-sample performance.
Both approaches have little bias despite estimation with only discrete-time
data where the order of moves is unobserved, but the variance reduction
from analytical gradients is substantial across all parameters.

With the larger sample size (Table~\ref{tab:mc:7x5:4000}), the differences
in standard errors between numerical and analytical gradients diminish
as expected, though analytical gradients still provide improved precision.
The convergence of both methods as the sample size increases confirms
the asymptotic properties discussed in Section~\ref{sec:loglik}.

\subsection{Computational Performance}

\begin{figure}[tbph]
  \begin{center}
    \includegraphics[width=\textwidth]{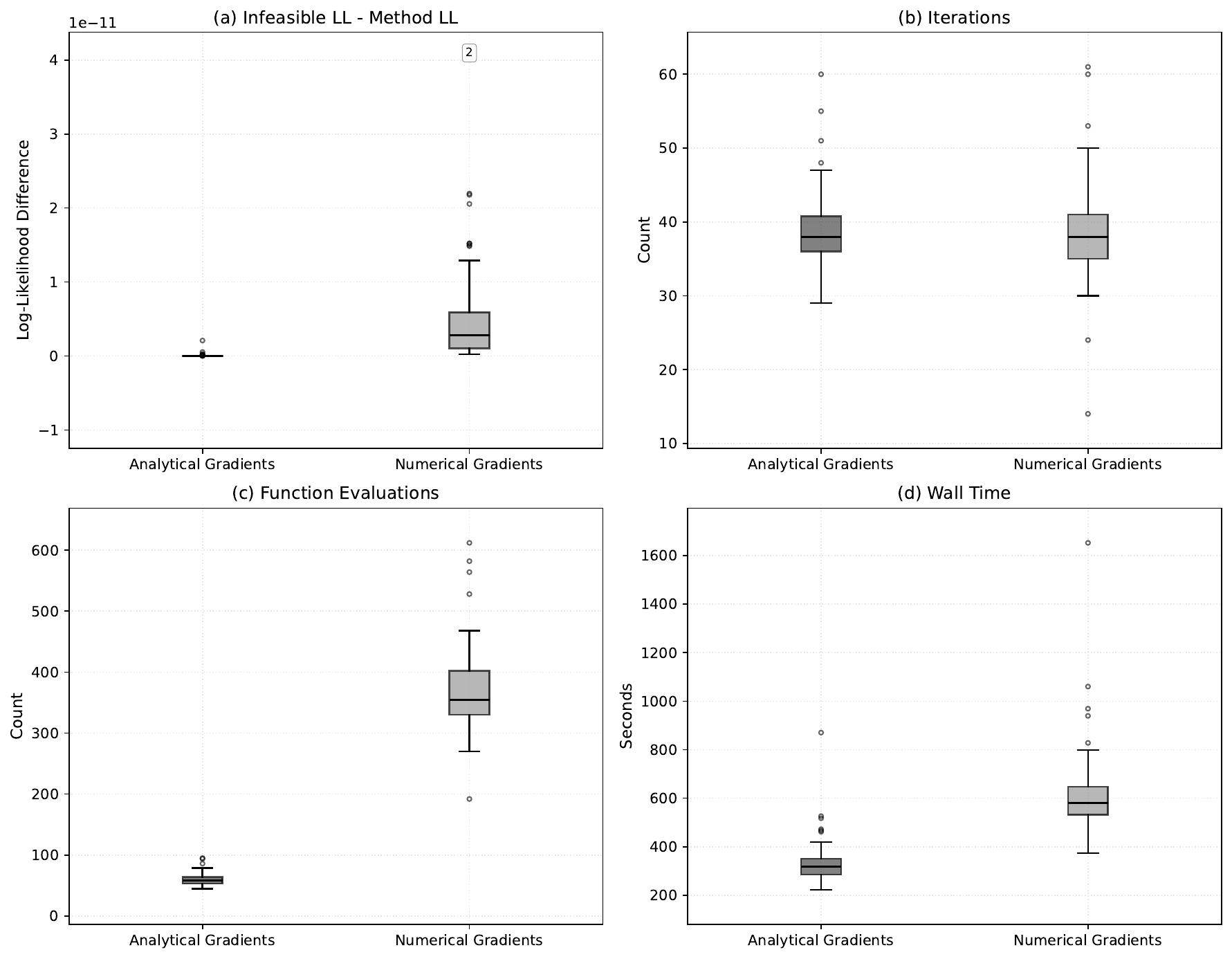}
    \caption{Computational performance comparison across
    Monte Carlo simulations (100 replications with 1,000 observations
    per replication).
    Panel (a) shows log-likelihood differences relative to the infeasible baseline.
    Panels (b)-(d) compare computational requirements: iterations,
    function evaluations, and wall time (seconds).
    Numbers above boxes indicate extreme outliers beyond the plot range,
    which were omitted for better scaling.}
    \label{fig:mc:computational_analysis:7x5:1000}
  \end{center}
\end{figure}

Beyond the statistical improvements, analytical gradients provide significant
computational efficiency gains.
Figure~\ref{fig:mc:computational_analysis:7x5:1000} provides a detailed
breakdown of computational performance across the Monte Carlo replications.
Panel (a) shows that using analytical gradients yields accurate
optimal log-likelihood values up to $10^{-11}$ precision, while numerical
gradients exhibit larger dispersion.
Panel (b) shows that both methods require approximately the same number
of iterations, but panels (c) and (d) reveal large differences in
the number of functional evaluations and computation times.

From Table~\ref{tab:mc:7x5:1000}, we see that using analytical
gradients reduces the average estimation time from 605.7 to 329.4
seconds---a 46\% reduction---and reduces the average number of
functional evaluations from 370.9 to 59.8---an 84\% reduction.
This is primarily due to the elimination of expensive
finite-difference gradient approximations at each iteration, which
involves solving the equilibrium of the model for each functional
evaluation.
The analytical derivatives are computed efficiently using
Algorithm~\ref{alg:expmvd} with implicit differentiation for the
equilibrium value functions.

\begin{table}[tbph]
\centering
\caption{Sparse Matrix Exponential Performance Comparison}
\label{tab:matrix_expm}
\begin{tabular}{lrrrrrrrr}
\toprule
 & \multicolumn{3}{c}{Full $\exp(Q)$} & \multicolumn{3}{c}{Columns of $\exp(Q)$} & \multicolumn{2}{c}{Difference} \\
\cmidrule(lr){2-4} \cmidrule(lr){5-7} \cmidrule(lr){8-9}
Model & Dense & Sparse & Speedup & Dense & Sparse & Speedup & Absolute & Relative \\
\midrule
2 $\times$ 2 & 0.24 & 3.73 & 0.06$\times$ & 0.22 & 3.76 & 0.06$\times$ & 6.0e-14 & 1.4e-12 \\
3 $\times$ 2 & 0.25 & 7.51 & 0.03$\times$ & 0.25 & 7.48 & 0.03$\times$ & 5.8e-14 & 4.4e-12 \\
4 $\times$ 2 & 0.33 & 15.47 & 0.02$\times$ & 0.31 & 15.45 & 0.02$\times$ & 1.5e-14 & 4.2e-12 \\
4 $\times$ 3 & 0.46 & 24.36 & 0.02$\times$ & 0.42 & 24.45 & 0.02$\times$ & 1.1e-14 & 7.8e-12 \\
5 $\times$ 3 & 1.54 & 50.76 & 0.03$\times$ & 1.50 & 50.93 & 0.03$\times$ & 9.4e-15 & 2.1e-11 \\
6 $\times$ 3 & 5.37 & 109.88 & 0.05$\times$ & 5.29 & 109.38 & 0.05$\times$ & 1.2e-14 & 9.7e-11 \\
6 $\times$ 4 & 13.06 & 160.11 & 0.08$\times$ & 12.76 & 123.53 & 0.10$\times$ & 1.9e-14 & 2.6e-10 \\
7 $\times$ 4 & 58.79 & 385.80 & 0.15$\times$ & 56.86 & 150.13 & 0.38$\times$ & 5.7e-15 & 2.3e-10 \\
7 $\times$ 5 & 97.23 & 534.95 & 0.18$\times$ & 96.45 & 166.10 & 0.58$\times$ & 1.4e-14 & 8.9e-10 \\
8 $\times$ 4 & 273.47 & 1,006.82 & 0.27$\times$ & 280.24 & 196.82 & 1.42$\times$ & 1.2e-14 & 1.7e-09 \\
8 $\times$ 5 & 480.16 & 1,397.82 & 0.34$\times$ & 465.67 & 218.51 & 2.13$\times$ & 6.8e-15 & 1.0e-09 \\
8 $\times$ 6 & 750.00 & 1,911.01 & 0.39$\times$ & 754.57 & 248.48 & 3.04$\times$ & 6.7e-15 & 1.3e-09 \\
9 $\times$ 5 & 2,818.66 & 4,510.21 & 0.62$\times$ & 2,824.37 & 348.67 & 8.10$\times$ & 1.1e-14 & 9.2e-09 \\
9 $\times$ 6 & 5,097.41 & 6,420.33 & 0.79$\times$ & 5,096.37 & 414.94 & 12.28$\times$ & 8.7e-15 & 6.5e-09 \\
10 $\times$ 6 & 32,933.93 & 25,448.87 & 1.29$\times$ & 32,933.97 & 813.56 & 40.48$\times$ & 1.3e-14 & 1.7e-08 \\
\bottomrule
\end{tabular}
\vspace{1ex}
\footnotesize
Note: Full = complete matrix exponential $\exp(Q)$, Columns = $\min\lbrace 200, K \rbrace$ columns of $\exp(Q)$,
Times in milliseconds. Speedup = Dense time / Sparse time ($>1$ means Sparse is faster).
\end{table}

Finally, to focus on the performance of the sparse matrix exponential
algorithm from Section~\ref{sec:loglik}, in Table~\ref{tab:matrix_expm}
we compare the standard dense matrix exponential function
\texttt{expm} from SciPy (based on Pad\'{e} approximation),
with the uniformization-based sparse version described in
Algorithm~\ref{alg:expmvd}.
First we compare computation of the full matrix exponential
$\exp(Q)$ for each model.
For smaller models, there is no advantage to using sparse
methods to compute the full matrix exponential, but for
the largest model we do begin to see a speedup.
For the models considered, the most significant
benefit of sparse methods occurs when we only need
to compute a relatively small number of columns, such as
when evaluating the log-likelihood function over a finite sample.
As we showed in Section~\ref{sec:loglik}, we only need to compute
columns corresponding to observed destination states,
which is typically much smaller than $K$ for moderate sample sizes.
Under the ``Columns of $\exp(Q)$'' heading in Table~\ref{tab:matrix_expm},
where we compute $\min\{200, K\}$ columns of $\exp(Q)$ (e.g., for a sample of 200 observations),
we see speedups of up to $40.48\times$.
Comparing the accuracy of the sparse method using the dense
method as the baseline, we see that even for very large
models the relative errors are all on the order of
$10^{-8}$ or smaller,
showing that the sparse uniformization-based method
is both efficient and accurate for maximum likelihood
estimation.

The choice between sparse and dense methods depends on both the state
space dimension $K$ and the number of matrix exponential columns required.
Dense methods compute the full $K \times K$ matrix exponential at cost
$O(K^3)$, making them efficient for small problems
or when many columns are needed.  Sparse methods avoid this cubic scaling by
computing individual columns at cost approximately $O(K)$ per column, becoming
advantageous when computing only $c$ columns with $c \ll K$.
In maximum likelihood estimation with sample size $T$, the number of required
columns is bounded by the number of unique destination states observed,
typically $O(\min\{T, K\})$.  Thus, sparse methods are recommended when
$T \ll K^2$, which encompasses most empirically relevant settings with
moderate to large state spaces.

\section{Conclusion}
\label{sec:conclusion}

This paper provides theoretical foundations and practical methods for
estimation of continuous-time dynamic discrete choice models, with
applications ranging from single-agent problems to multi-agent games.
Our contributions address both econometric and computational
challenges that have limited the adoption of continuous-time methods
despite their theoretical advantages.

From an econometric perspective, we provide formal convergence
guarantees for value iteration with fixed beliefs in continuous-time
settings and develop a complete analytical derivative chain for
maximum likelihood estimation.
These analytical derivatives are not merely a computational
convenience---they eliminate approximation errors from numerical
gradients that affect finite-sample performance, as our Monte Carlo
experiments demonstrate with up to eight-fold reductions in parameter
standard errors.
This has significant implications for empirical researchers working
with real-world datasets.

From a computational perspective, our application of uniformization
exploits the sparse structure inherent in many economic models to
achieve substantial efficiency gains.
Specifically, we have (1) established theoretical convergence
properties for value iteration with fixed beliefs, (2) developed a
polyalgorithm combining value iteration with Newton-Kantorovich
methods for robust equilibrium computation, (3) introduced a
uniformization-based representation that provides direct analogies
with discrete-time models, and (4) demonstrated efficient estimation
with discrete-time snapshot data using analytical derivatives of
sparse matrix exponentials.
Our experiments show that sparse methods achieve speedups of up to
40$\times$ when computing only the columns needed for likelihood
evaluation, while analytical gradients reduce computation time by
46\%.

By reducing computational barriers and improving finite sample
statistical properties, these methods enable researchers to estimate
richer, more realistic models that were previously infeasible.  The
techniques apply to other models involving finite-state
continuous-time Markov chains, including single-agent dynamic discrete
choice, duration models with time-varying covariates, and
regime-switching models.  This opens new possibilities for empirical
work in industrial organization, labor economics, finance, and related
fields.

\appendix

\section{Proofs}
\label{sec:proofs}

\subsection{Proof of Theorem \ref{thm:vi:non-uniform}}

Let $V_i, V_i' \in \R^K$ denote two arbitrary value functions for player $i$.
Then, from \eqref{eq:bellman} we have
\begin{align*}
  &\norm{T_{i}^{\varsigma_i} V_i - T_{i}^{\varsigma_i} V_i'}_\infty\\
  &= \max_k \abs{T_{i}^{\varsigma_i} V_{ik} - T_{i}^{\varsigma_i} V_{ik}'} \\
  &= \max_k \biggl\lvert
    \frac{u_{ik} +
    \sum_{k' \neq k} q_{0kk'} V_{ik'} +
    \sum_{m \neq i} \lambda_{mk} \sum_j \varsigma_{imjk} V_{i,l(m,j,k)} +
    \lambda_{ik} \E \max_j \{ \psi_{ijk} + \varepsilon_{ijk} + V_{i,l(i,j,k)} \}}
    {\rho_i + \unifrate_k} \\
  &\qquad -
    \frac{u_{ik} +
    \sum_{k' \neq k} q_{0kk'} V_{ik'}' +
    \sum_{m \neq i} \lambda_{mk} \sum_j \varsigma_{imjk} V_{i,l(m,j,k)}' +
    \lambda_{ik} \E \max_j \{ \psi_{ijk} + \varepsilon_{ijk} + V_{i,l(i,j,k)}' \}}
    {\rho_i + \unifrate_k}
    \biggr\rvert \\
  &= \max_k \biggl\lvert \frac{1}{\rho_i + \unifrate_k} \left( \sum_{k' \neq k} q_{0kk'} (V_{ik'} - V_{ik'}')
    + \sum_{m \neq i} \lambda_{mk} \sum_j \varsigma_{imjk} (V_{i,l(m,j,k)} - V_{i,l(m,j,k)}') \right. \\
  &\qquad \left. +\lambda_{ik} \left[ \E \max_j \left\{ \psi_{ijk} + \varepsilon_{ijk} + V_{i,l(i,j,k)} \right\}
                                    - \E \max_j \left\{ \psi_{ijk} + \varepsilon_{ijk} + V_{i,l(i,j,k)}' \right\}\right]
          \right) \biggr\rvert \\
  &\leq \max_k \frac{1}{\rho_i + \unifrate_k}
      \left( \sum_{k' \neq k} q_{0kk'} \abs{V_{ik'} - V_{ik'}'}
          + \sum_{m \neq i} \lambda_{mk} \sum_j \varsigma_{imjk} \abs{V_{i,l(m,j,k)} - V_{i,l(m,j,k)}'}
          \right. \\
  &\qquad \left.
          + \lambda_{ik} \max_j \abs{ V_{i,l(i,j,k)} - V_{i,l(i,j,k)}' } \right) \\
  &\leq \max_k \left[ \frac{\unifrate_k}{\rho_i + \unifrate_k} \right] \max_{k'} \abs{V_{ik'} - V_{ik'}'} \\
  &= \beta_i \norm{V_{i} - V_{i}'}_\infty
\end{align*}
The first equality follows by definition of supremum norm with
finite $\intstatespace$, the second from the definition of the Bellman operator in
\eqref{eq:bellman}, and the third from collecting a common
denominator, noting that $u_{ik}$ cancels out, and then collecting
terms.

The first inequality follows from the triangle inequality, but the final term
requires further explanation:
We use linearity of the expectation and the fact that
$\abs{ \max_j a_j - \max_j b_j } \leq \max_j \abs{ a_j - b_j }$
for two vectors $a$ and $b$.
Additionally, after subtracting and canceling the
$\psi_{ijk} + \varepsilon_{ijk}$ terms, the resulting
expression is constant with respect to the random variables $\varepsilon_{ijk}$,
so the expectation can be dropped:
\begin{align*}
  & \abs{\E \max_j \left\{ \psi_{ijk} + \varepsilon_{ijk} + V_{i,l(i,j,k)} \right\} -
    \E \max_j \left\{ \psi_{ijk} + \varepsilon_{ijk} + V_{i,l(i,j,k)}' \right\}} \\
  &\quad\leq
    \E \left[ \max_j \abs{ \left( \psi_{ijk} + \varepsilon_{ijk} + V_{i,l(i,j,k)} \right) -
    \left( \psi_{ijk} + \varepsilon_{ijk} + V_{i,l(i,j,k)}' \right) } \right] \\
  &\quad=
    \E \left[ \max_j \abs{ V_{i,l(i,j,k)} - V_{i,l(i,j,k)}' } \right] \\
  &\quad=
  \max_j \abs{ V_{i,l(i,j,k)} - V_{i,l(i,j,k)}' } \\
  &\quad\leq
  \max_{k'} \abs{ V_{ik'} - V_{ik'}' }
\end{align*}
The last line here follows from $\max_j \abs{ V_{i,l(i,j,k)} - V_{i,l(i,j,k)}' } \leq \max_{k'} \abs{ V_{ik'} - V_{ik'}' }$,
since for each $j$, the continuation state $l(i,j,k)$ is in $\intstatespace$,
so the maximum over choices $j$ is bounded by the maximum over all states $k'$.

This leads us to the second inequality in the main derivation after noting that the beliefs are each probabilities in $[0,1]$,
and using the definition of $\unifrate_k$ from \eqref{eq:unifrate_k}.
The final inequality follows from recognizing that the maximum over $k$ of the fraction
$\frac{\unifrate_k}{\rho_i + \unifrate_k}$
is precisely $\beta_i$ as defined in \eqref{eq:beta_i}.

\subsection{Proof of Theorem~\ref{thm:pe:uniform}}

Recall the Bellman equation from \eqref{eq:bellman}.
Multiplying both sides by the sum of rates in the denominator yields
\begin{multline*}
  \left[ \rho_i + \unifrate_k \right] V_{ik}
  = u_{ik} + \sum_{k' \neq k} q_{0kk'} V_{ik'} +
  \sum_{m \neq i} \lambda_{mk} \sum_j \varsigma_{imjk} V_{i,l(m,j,k)} \\ +
  \lambda_{ik} \E \max_j \{ \psi_{ijk} + \varepsilon_{ijk} + V_{i,l(i,j,k)} \}.
\end{multline*}
Collecting terms yields a representation of the instantaneous
increment to the value function:
\begin{multline*}
  \rho_i V_{ik}
  = u_{ik} + \sum_{k' \neq k} q_{0kk'} \left( V_{ik'} - V_{ik} \right) +
  \sum_{m \neq i} \lambda_{mk} \sum_j \varsigma_{imjk} \left( V_{i,l(m,j,k)} - V_{ik} \right) \\ +
  \lambda_{ik} \E \max_j \{ \psi_{ijk} + \varepsilon_{ijk} + V_{i,l(i,j,k)} - V_{ik} \}.
\end{multline*}

First, recall that since the rows of $Q_0$ sum to zero, we have
$q_{0kk} = -\sum_{k' \neq k} q_{0kk'}$.
Therefore,
\begin{equation*}
  \sum_{k' \neq k} q_{0kk'} \left( V_{ik'} - V_{ik} \right)
  = \sum_{k' \neq k} q_{0kk'} V_{ik'} - \sum_{k' \neq k} q_{0kk'} V_{ik}
  = \sum_{k' \neq k} q_{0kk'} V_{ik'} + q_{0kk} V_{ik}
  = \sum_{k'=1}^K q_{0kk'} V_{ik'}
\end{equation*}
which is the $k$-th row of $Q_0$ multiplied by the column vector $V_i$ (i.e., the $k$-th element of $Q_0 V_i$).

Next, consider the term for rival firm $m$. Recall that the choice probabilities
sum to one across choices $j$ and that choice $j=0$ is a continuation choice
such that $l(m,0,k) = k$:
\begin{align*}
  \lambda_{mk} \sum_j \varsigma_{imjk} \left( V_{i,l(m,j,k)} - V_{ik} \right)
  &= \lambda_{mk} \sum_{j > 0} \varsigma_{imjk} V_{i,l(m,j,k)} + \lambda_{mk} \varsigma_{im0k} V_{ik} - \lambda_{mk} V_{ik}\\
  &= \lambda_{mk} \sum_{j > 0} \varsigma_{imjk} V_{i,l(m,j,k)} - \lambda_{mk} \sum_{j > 0} \varsigma_{imjk} V_{ik}\\
  &= \sum_{k' = 1}^K q_{mkk'} V_{ik'},
\end{align*}
which is the $k$-th row of $Q_m$ multiplied by the column vector $V_i$ (i.e., the $k$-th element of $Q_m V_i$).

Next, we show that we can write the expectation of the maximum term in
terms of choice probabilities, following \cite*{abbe-2016} and
\cite{aguirregabiria-mira-2002,aguirregabiria-mira-2007}.
Using the definition of $C_i(\sigma_i)$ from the statement of the theorem,
note that we can write the final term, related to the agent's own
optimization, as
\begin{equation*}
  \lambda_{ik} \E \max_j \{ \psi_{ijk} + \varepsilon_{ijk} + V_{i,l(i,j,k)} - V_{ik} \}
  = \lambda_{ik} C_{ik}(\sigma_i) + \sum_{k'=1}^K q_{ikk'} V_{ik'}.
\end{equation*}
Note that the second term is the product of the $k$-th row of $Q_i$ with the vector $V_i$.

Combining these results, we can write the vectorized
value function as
\begin{equation*}
  \rho_i V_i = u_i + Q_0 V_i + \sum_{m \neq i} Q_m V_i + L_i C_i(\sigma_i) + Q_i V_i.
\end{equation*}
Noting that $Q = Q_0 + \sum_{m=1}^{\Nplayers} Q_m$, with each matrix depending on the
$\sigma$ for $m > 1$,
we can write the vectorized form of the
value function more simply as
\begin{equation}
  \label{eq:instantaneous}
  \rho_i V_i = u_i + L_i C_i(\sigma_i) + Q V_i.
\end{equation}

Since $\unifrate$ satisfies \eqref{eq:unifrate} and is a valid uniformization rate,
we may write $Q = \unifrate \left( \Sigma(\sigma) - I \right)$
where $\Sigma(\sigma)$ is the stochastic transition matrix that depends on the policies through the $Q_i$ matrices.

Finally, to derive the stated uniform representation of $V_i$, note that from
\eqref{eq:instantaneous} we have
\begin{equation*}
  \rho_i V_i = u_i + L_i C_i(\sigma_i) + \unifrate \left( \Sigma(\sigma) - I \right) V_i
\end{equation*}
and therefore,
\begin{align*}
  V_i &= \frac{1}{\rho_i + \unifrate} [ u_i + L_i C_i(\sigma_i) ] + \frac{\unifrate}{\rho_i + \unifrate} \Sigma(\sigma) V_i \\
  &\equiv U_i(\sigma) + \overline\beta_i \Sigma(\sigma) V_i
\end{align*}
where $U_i(\sigma)$ and $\overline\beta_i$ are defined in
the statement of the theorem.

To show that $\overline{\Gamma}_{i}^{\sigma}$ is a contraction with modulus $\overline\beta_i$,
note that for any $V_i, V_i' \in \mathbb{R}^K$:
\begin{align*}
\|\overline{\Gamma}_{i}^{\sigma} V_i - \overline{\Gamma}_{i}^{\sigma} V_i'\|_\infty
&= \|U_i(\sigma) + \overline\beta_i \Sigma(\sigma) V_i - U_i(\sigma) - \overline\beta_i \Sigma(\sigma) V_i'\|_\infty \\
&= \overline\beta_i \|\Sigma(\sigma) (V_i - V_i')\|_\infty \\
&\leq \overline\beta_i \|V_i - V_i'\|_\infty
\end{align*}
The last inequality follows because $\Sigma(\sigma)$ is a stochastic matrix with row sums equal to one.
Since $\overline\beta_i \in (0,1)$, this establishes that $\overline{\Gamma}_{i}^{\sigma}$ is a contraction mapping.

\subsection{Proof of Lemma~\ref{lem:contraction_moduli}}

Recall that $\unifrate_k = \sum_{k' \neq k} q_{0kk'} + \sum_m \lambda_{mk}$ denotes the maximum exit rate from state $k$ from \eqref{eq:unifrate_k}.
From the definitions in Theorems~\ref{thm:vi:non-uniform} and \ref{thm:pe:uniform}, we have:
\begin{equation*}
\beta_i = \max_k \frac{\unifrate_k}{\rho_i + \unifrate_k} \quad \text{and} \quad
\overline{\beta}_i = \frac{\unifrate}{\rho_i + \unifrate}.
\end{equation*}
Since $\unifrate \geq \max_k \unifrate_k$ by \eqref{eq:unifrate} and the function $f(x) = \frac{x}{\rho_i + x}$ is increasing in $x$ for $x > 0$, we have
\begin{equation*}
\overline{\beta}_i = f(\unifrate) \geq f\left(\max_k \unifrate_k\right) = \max_k f(\unifrate_k) = \beta_i,
\end{equation*}
where the second equality holds because $f$ is increasing.
Therefore $\beta_i \leq \overline{\beta}_i$, with equality when $\unifrate_k$ is constant across all states $k$
and $\unifrate = \max_k \unifrate_k$.

\section{Structure of the Equilibrium System Jacobian}
\label{sec:jacobian_structure}

This section provides explicit formulas for computing the Jacobian
$\frac{\partial T}{\partial V}$ of the equilibrium system operator.
These derivatives are used in the Newton-Kantorovich iterations of the
polyalgorithm and for computing derivatives of the value function
with respect to parameters via implicit differentiation.

Recall that the value function vector $V$ is stacked across players and states:
$V = (V_{11}, \ldots, V_{\Nplayers K})^\top$.
The equilibrium system operator $T(V)$ is defined in \eqref{eq:equilibrium_system}
in terms of the operator $T_{i}^{\varsigma_i}$, defined by the Bellman equation in \eqref{eq:bellman},
with beliefs that satisfy
$\varsigma_{imjk} = \sigma_{mjk} = \Pr[j \in \argmax_{j'} \{\psi_{mj'k} + \varepsilon_{mj'k} + V_{m,l(m,j',k)}\}]$
for all players $i$, rival players $m$, choices $j$, and states $k$.
This belief consistency condition means that $T_{ik}$ depends on $V_m$ for all $m$ both through
player $i$'s own value function and through beliefs about rivals.
The Jacobian $\frac{\partial T}{\partial V}$ has elements
$\frac{\partial T_{ik}}{\partial V_{mk'}}$
where $(i,k)$ indexes the row (player $i$ in state $k$)
and $(m,k')$ indexes the column (derivative with respect to player $m$'s value in state $k'$).

We analyze the Jacobian elements in two parts: derivatives with respect to
player $i$'s own value function, and derivatives with respect to rival players'
value functions.

Consider the derivatives with respect to player $i$'s own value function.
By the Williams-Daly-Zachary theorem \citep[][Theorem 3.1]{rust-1994-structural},
the derivative of the $\E\max$ term with respect to continuation values
equals the choice probabilities.
This yields:
\begin{equation*}
  \frac{\partial T_{ik}}{\partial V_{ik'}} = \frac{1}{\rho_i + \unifrate_k} \left[
    q_{0kk'} \1\{k' \neq k\} +
    \sum_{m \neq i} \lambda_{mk} \sum_j \sigma_{mjk} \1\{l(m,j,k) = k'\} +
    \lambda_{ik} \sum_j \sigma_{ijk} \1\{l(i,j,k) = k'\}
  \right].
\end{equation*}

For the diagonal element, Assumption~\ref{assp:dn} gives $l(i,0,k) = k$, so:
\begin{equation*}
  \frac{\partial T_{ik}}{\partial V_{ik}} = \frac{1}{\rho_i + \unifrate_k} \left[
    \sum_{m \neq i} \lambda_{mk} \sum_j \sigma_{mjk} \1\{l(m,j,k) = k\} +
    \lambda_{ik} \sigma_{i0k}
  \right].
\end{equation*}
Recognizing that $\sigma_{i0k} = 1 - \sigma_{i1k}$ for the binary action case
and more generally $\sum_j \sigma_{mjk} \1\{l(m,j,k) = k\} = 1 - \sigma_{m1k}$
when action $j=1$ is the only switching action, this can be written as:
\begin{equation*}
  \frac{\partial T_{ik}}{\partial V_{ik}} = \frac{1}{\rho_i + \unifrate_k} \left[
    \sum_{m \neq i} \lambda_{mk} (1 - \sigma_{m1k}) +
    \lambda_{ik} (1 - \sigma_{i1k})
  \right].
\end{equation*}
The diagonal element $\frac{\partial T_{ik}}{\partial V_{ik}}$
includes contributions from \emph{all} players' continuation
probabilities, not just player $i$'s own continuation probability.
The term $\lambda_{mk}(1-\sigma_{m1k})$ appears because player $i$'s value at state $k$
depends on what happens when each rival player $m$ makes a choice.
If player $m$ chooses to continue (probability $1-\sigma_{m1k}$), the state remains
unchanged at $k$ and player $i$'s continuation value is $V_{ik}$.
For off-diagonal elements where $k' \neq k$:
\begin{equation*}
  \frac{\partial T_{ik}}{\partial V_{ik'}}
  = \frac{1}{\rho_i + \unifrate_k} \left[
    q_{0kk'} +
    \sum_{m \neq i} \lambda_{mk} \sum_j \sigma_{mjk} \1\{l(m,j,k) = k'\} +
    \lambda_{ik} \sum_j \sigma_{ijk} \1\{l(i,j,k) = k'\}
  \right].
\end{equation*}

For $m \neq i$, the operator $T_{ik}$ depends on $V_m$ through the belief consistency condition
$\varsigma_{mjk} = \sigma_{mjk}(V_m)$.
From the Bellman equation \eqref{eq:bellman}, the numerator includes the term
$\lambda_{mk} \sum_j \sigma_{mjk}(V_m) V_{i,l(m,j,k)},$
which depends on $V_m$ through the choice probabilities.
The denominator $\rho_i + \unifrate_k$ is constant, independent of choice probabilities.
Therefore:
\begin{equation*}
  \frac{\partial T_{ik}}{\partial V_{mk'}} = \frac{\lambda_{mk}}{\rho_i + \unifrate_k} \sum_j \frac{\partial \sigma_{mjk}}{\partial V_{mk'}} V_{i,l(m,j,k)}.
\end{equation*}
For the binary action case with $j \in \{0,1\}$ where action $j=0$ continues
at state $k$ and action $j=1$ switches to state $l(m,1,k)$,
this sum expands as:
\begin{equation*}
  \frac{\partial T_{ik}}{\partial V_{mk'}} = \frac{\lambda_{mk}}{\rho_i + \unifrate_k} \left[
    \frac{\partial \sigma_{m0k}}{\partial V_{mk'}} V_{ik} +
    \frac{\partial \sigma_{m1k}}{\partial V_{mk'}} V_{i,l(m,1,k)}
  \right].
\end{equation*}
Using $\sigma_{m0k} = 1 - \sigma_{m1k}$ and the logit derivatives given in
Section~\ref{sec:loglik},
the sum simplifies to a difference in values:
\begin{equation*}
  \frac{\partial T_{ik}}{\partial V_{mk'}} = \frac{\lambda_{mk}}{\rho_i + \unifrate_k}
    \frac{\partial \sigma_{m1k}}{\partial V_{mk'}}
    \left(V_{i,l(m,1,k)} - V_{ik}\right).
\end{equation*}

\bibliographystyle{chicago}
\bibliography{ctcomp}

\end{document}

%% file: 2x2x2-entry-standard.tex
\begin{tikzpicture}[->, >=Stealth, auto, node distance=3cm, baseline=(current bounding box.center)]
\tikzstyle{every state}=[circle, draw=stateBorderColor, fill=stateColor, thick, text=black, minimum size=0.9cm, font=\small]
\tikzstyle{group box}=[draw=black, thick, rounded corners, fill=groupColor]

\node[state] (L00) {$(\text{L},0,0)$};
\node[state, right of=L00, node distance=2.8cm] (L10) {$(\text{L},1,0)$};
\node[state, below of=L00, node distance=2.8cm] (L01) {$(\text{L},0,1)$};
\node[state, right of=L01, node distance=2.8cm] (L11) {$(\text{L},1,1)$};

\begin{pgfonlayer}{background}
\node[group box, fit=(L00)(L10)(L01)(L11), inner sep=30pt, label=above:{Demand = L}] (Lbox) {};
\end{pgfonlayer}

\node[state, right of=L10, node distance=5cm]   (H00) {$(\text{H},0,0)$};
\node[state, right of=H00, node distance=2.8cm] (H10) {$(\text{H},1,0)$};
\node[state, below of=H00, node distance=2.8cm] (H01) {$(\text{H},0,1)$};
\node[state, right of=H01, node distance=2.8cm] (H11) {$(\text{H},1,1)$};

\begin{pgfonlayer}{background}
\node[group box, fit=(H00)(H10)(H01)(H11), inner sep=30pt, label=above:{Demand = H}] (Hbox) {};
\end{pgfonlayer}

\path[firmOneColor]
(L00) edge[bend left=15] node[above] {$\lambda \sigma_{111}$} (L10)
(L10) edge[bend left=15] node[below] {$\lambda \sigma_{112}$} (L00)
(L01) edge[bend left=15] node[above] {$\lambda \sigma_{113}$} (L11)
(L11) edge[bend left=15] node[below] {$\lambda \sigma_{114}$} (L01);

\path[firmTwoColor]
(L00) edge[bend left=15] node[right] {$\lambda \sigma_{211}$} (L01)
(L01) edge[bend left=15] node[left] {$\lambda \sigma_{213}$} (L00)
(L10) edge[bend left=15] node[right] {$\lambda \sigma_{212}$} (L11)
(L11) edge[bend left=15] node[left] {$\lambda \sigma_{214}$} (L10);

\path[firmOneColor]
(H00) edge[bend left=15] node[above] {$\lambda \sigma_{115}$} (H10)
(H10) edge[bend left=15] node[below] {$\lambda \sigma_{116}$} (H00)
(H01) edge[bend left=15] node[above] {$\lambda \sigma_{117}$} (H11)
(H11) edge[bend left=15] node[below] {$\lambda \sigma_{118}$} (H01);

\path[firmTwoColor]
(H00) edge[bend left=15] node[right] {$\lambda \sigma_{215}$} (H01)
(H01) edge[bend left=15] node[left] {$\lambda \sigma_{217}$} (H00)
(H10) edge[bend left=15] node[right] {$\lambda \sigma_{216}$} (H11)
(H11) edge[bend left=15] node[left] {$\lambda \sigma_{218}$} (H10);

\path[demandColor, thick]
(Lbox.east)++(0,+0.5) edge[bend left=25] node[above] {$\gamma$} (Hbox.west)
(Hbox.west)++(0,-0.5) edge[bend left=25] node[below] {$\gamma$} (Lbox.east);


\end{tikzpicture}

%% file: 2x2x2-entry-uniform.tex
\begin{tikzpicture}[->, >=Stealth, auto, node distance=3cm, baseline=(current bounding box.center)]
\tikzstyle{every state}=[circle, draw=stateBorderColor, fill=stateColor, thick, text=black, minimum size=0.9cm, font=\small]
\tikzstyle{group box}=[draw=black, thick, rounded corners, fill=groupColor]

\node[state] (L00) {$(\text{L},0,0)$};
\node[state, right of=L00, node distance=2.8cm] (L10) {$(\text{L},1,0)$};
\node[state, below of=L00, node distance=2.8cm] (L01) {$(\text{L},0,1)$};
\node[state, right of=L01, node distance=2.8cm] (L11) {$(\text{L},1,1)$};

\begin{pgfonlayer}{background}
\node[group box, fit=(L00)(L10)(L01)(L11), inner sep=30pt, label=above:{Demand = L}] (Lbox) {};
\end{pgfonlayer}

\node[state, right of=L10, node distance=5cm]   (H00) {$(\text{H},0,0)$};
\node[state, right of=H00, node distance=2.8cm] (H10) {$(\text{H},1,0)$};
\node[state, below of=H00, node distance=2.8cm] (H01) {$(\text{H},0,1)$};
\node[state, right of=H01, node distance=2.8cm] (H11) {$(\text{H},1,1)$};

\begin{pgfonlayer}{background}
\node[group box, fit=(H00)(H10)(H01)(H11), inner sep=30pt, label=above:{Demand = H}] (Hbox) {};
\end{pgfonlayer}


\path
(L00) edge[loop, in=120, out=150, looseness=5] node[left]  {$\Sigma_{11}$} (L00)
(L10) edge[loop, in=30,  out=60,  looseness=5] node[above] {$\Sigma_{22}$} (L10)
(L01) edge[loop, in=210, out=240, looseness=5] node[below] {$\Sigma_{33}$} (L01)
(L11) edge[loop, in=300, out=330, looseness=5] node[right] {$\Sigma_{44}$} (L11)
(H00) edge[loop, in=120, out=150, looseness=5] node[left]  {$\Sigma_{55}$} (H00)
(H10) edge[loop, in=30,  out=60,  looseness=5] node[above] {$\Sigma_{66}$} (H10)
(H01) edge[loop, in=210, out=240, looseness=5] node[below] {$\Sigma_{77}$} (H01)
(H11) edge[loop, in=300, out=330, looseness=5] node[right] {$\Sigma_{88}$} (H11);

\path[firmOneColor]
(L00) edge[bend left=15] node[above] {$\frac{\lambda \sigma_{111}}{\unifrate}$} (L10) 
(L10) edge[bend left=15] node[below] {$\frac{\lambda \sigma_{112}}{\unifrate}$} (L00) 
(L01) edge[bend left=15] node[above] {$\frac{\lambda \sigma_{113}}{\unifrate}$} (L11) 
(L11) edge[bend left=15] node[below] {$\frac{\lambda \sigma_{114}}{\unifrate}$} (L01) 
(H00) edge[bend left=15] node[above] {$\frac{\lambda \sigma_{115}}{\unifrate}$} (H10) 
(H10) edge[bend left=15] node[below] {$\frac{\lambda \sigma_{116}}{\unifrate}$} (H00) 
(H01) edge[bend left=15] node[above] {$\frac{\lambda \sigma_{117}}{\unifrate}$} (H11) 
(H11) edge[bend left=15] node[below] {$\frac{\lambda \sigma_{118}}{\unifrate}$} (H01); 

\path[firmTwoColor]
(L00) edge[bend left=15] node[right] {$\frac{\lambda \sigma_{211}}{\unifrate}$} (L01) 
(L01) edge[bend left=15] node[left]  {$\frac{\lambda \sigma_{213}}{\unifrate}$} (L00) 
(L10) edge[bend left=15] node[right] {$\frac{\lambda \sigma_{212}}{\unifrate}$} (L11) 
(L11) edge[bend left=15] node[left]  {$\frac{\lambda \sigma_{214}}{\unifrate}$} (L10) 
(H00) edge[bend left=15] node[right] {$\frac{\lambda \sigma_{215}}{\unifrate}$} (H01) 
(H01) edge[bend left=15] node[left]  {$\frac{\lambda \sigma_{217}}{\unifrate}$} (H00) 
(H10) edge[bend left=15] node[right] {$\frac{\lambda \sigma_{216}}{\unifrate}$} (H11) 
(H11) edge[bend left=15] node[left]  {$\frac{\lambda \sigma_{218}}{\unifrate}$} (H10); 

\path[demandColor, thick]
(Lbox.east)++(0,+0.5) edge[bend left=25] node[above] {$\frac{\gamma}{\unifrate}$} (Hbox.west)
(Hbox.west)++(0,-0.5) edge[bend left=25] node[below] {$\frac{\gamma}{\unifrate}$} (Lbox.east);


\end{tikzpicture}